\documentclass{iopart}
\usepackage{bm}
\usepackage{hyperref}

\usepackage{amssymb}
\usepackage{amsthm}
\usepackage{wasysym}

\theoremstyle{plain}

\font\SYM=msbm10 
\newcommand{\Real}{\mbox{\SYM R}}

\newcommand{\Sphere}{\mbox{\SYM S}}

\newcommand{\be}{\begin{equation}}
\newcommand{\ee}{\end{equation}}
\newcommand{\bea}{\begin{eqnarray}}
\newcommand{\eea}{\end{eqnarray}}

\newcommand{\beq}{\begin{equation}}
\newcommand{\eeq}{\end{equation}}

\font\tenscr=rsfs10 scaled1100
\font\sevenscr=rsfs7 
\font\fivescr=rsfs5 
\skewchar\tenscr='177
\skewchar\sevenscr='177
\skewchar\fivescr='177
\newfam\scrfam
\textfont\scrfam=\tenscr
\scriptfont\scrfam=\sevenscr
\scriptscriptfont\scrfam=\fivescr

\def\scri{{\fam\scrfam I}}

\newcommand*{\MEU}{Laboratoire Univers et Th\'eories, UMR 8102
  du C.N.R.S., Observatoire de Paris, Universit\'e Paris Diderot, F-92190 Meudon, France}
\newcommand*{\IAA}{Instituto de Astrof\'{\i}sica de Andaluc\'{\i}a,
  CSIC, Apartado Postal 3004, Granada 
        18080, Spain} 
\newcommand*{\QMWC}{School of Mathematical Sciences,
  Queen Mary, University of London,
  Mile End Road
  London E1 4NS
  U.K. } 
\begin{document}

\title[From Geometry to Numerics]{From Geometry to Numerics: interdisciplinary aspects in mathematical and numerical relativity}

\author{
Jos\'e Luis Jaramillo$^{1,3}$\footnote{Email address: jarama@iaa.es} 
Juan Antonio Valiente Kroon$^2$\footnote{Email address: j.a.valiente-kroon@qmul.ac.uk } 
and Eric Gourgoulhon$^3$ \footnote{Email address: eric.gourgoulhon@obspm.fr}}

\address{$^1$ \IAA}
\address{$^2$ \QMWC}
\address{$^3$ \MEU}

\date{12 December 2007}

\begin{abstract}
  This article reviews some aspects in the current relationship between
  mathematical and numerical General Relativity. Focus is placed on
  the description of isolated systems, with a particular emphasis on
  recent developments in the study of black holes. Ideas concerning
  asymptotic flatness, the initial value problem, the constraint
  equations, evolution formalisms, geometric inequalities and
  quasi-local black hole horizons are discussed on the light of the
  interaction between numerical and mathematical relativists.
\end{abstract}


\maketitle

\renewcommand{\thefootnote}{\arabic{footnote}}
\setcounter{footnote}{0}

\section{Introduction}
\label{s:Intro}

\subsection{Objectives}
In this review article we focus on specific problems which are of
interest both for numerical relativists and for geometers. A number of
review articles has been devoted in the last years to the technical
aspects and {\it state of the art} of each respective domain ---e.g.
Andersson \cite{And06}, Ashtekar \& Krishnan \cite{AshKri04}, Bartnik
\& Isenberg \cite{BarIse04}, Baumgarte \& Shapiro \cite{BauSha03},
Berger \cite{Ber02}, Booth \cite{Boo05}, Cook \cite{Coo00}, Friedrich \cite{Fri05},
Friedrich \& Rendall \cite{FriRen00}, Gourgoulhon
\cite{Gou07a,Gou07b}, Gundlach \& Mart\'{\i}n-Garc\'{\i}a \cite{GunMar07},
Klainerman \& Nicol\`o \cite{KlaNic99}, Krishnan \cite{Kri07}, 
Lehner \cite{Leh01} , Lehner \& Reula \cite{LehReu04}, 
Rendall \cite{Ren05}, Reula \cite{Reu98},
Shinkai \& Yoneda \cite{ShiYon02}, Thornburg \cite{Tho07}, Winicour
\cite{Win05}, and also \cite{NFNR07}. More theme specific reviews will
be referred to in the corresponding sections.  The aim here is not the
exhaustive description of the respective specific tools, but rather to
identify and discuss conceptual or structural challenges in General
Relativity which represent good candidates for a close collaboration
between geometers and numerical relativists.

This review is inspired in a workshop with a similar name that took
place in the context of the \emph{General Relativity Trimester} held
at the \emph{Institut Henri Poincar\'e}, Paris, from November 20 to
24, 2006 \cite{IHP}, which brought together specialists from both the
mathematical and numerical ends of General Relativity (GR). Due to the
broad range of topics covered in the workshop we have taken the
\emph{methodological} decision to restrict our attention to numerical
and mathematical aspects of the description of isolated systems in GR.
This somehow reflects the underlying ultimate interests of the
organisers, biased towards astrophysically motivated problems not
including Cosmology. Unavoidably, not all the topics discussed at the
time are covered in this review and on the other hand, some of the topics
discussed in the review did not have a counterpart talk in the
workshop.

\subsection{General thoughts}

\subsubsection{Why from Geometry to Numerics?} General Relativity is a
theory that describes the gravitational interaction as a manifestation of the
geometry of spacetime. It is therefore natural to expect that a
geometrical perspective may be helpful, and even sometimes fundamental, for
the further development of the theory at both conceptual and practical levels
---the latter understood as the explicit construction of solutions. Does this
point of view live up to the expectations raised?

In our opinion, the answer to this last question has two facets. On
the one hand, as discussed in \cite{Fri05}, understanding GR can be
seen as tantamount to understanding the properties of its solutions,
and in this sense \emph{``getting qualitative and quantitative
  (respectively analytical, theoretical, and numerical) control on the
  long time evolution of gravitational fields under general
  assumptions is still the most important open problem in classical
  general relativity''}. In this context there seems to be plenty of
occasion for the successful interaction between geometry and numerics,
not only in the direction suggested by the title of the review but also
on the converse one, from \emph{numerics to geometry}.

Regarding the second facet, arguably, the other great challenge for
the theory is to make full contact with observations ---in particular,
nowadays, in the area of gravitational waves physics. The nature of
this specific endeavour requires the efficient calculation of
spacetimes describing isolated systems.  This calculation, to be
physically realistic has to be performed using numerical simulations
of some particular formulation of the Einstein equations. In this
context, the fruitfulness of using geometry as a guideline in
numerical applications seems much harder to assess, since the
understanding of a particular analytical aspect of a geometric problem
does not ensure its successful numerical implementation. Furthermore,
it should not be forgotten that one is talking about the (potential)
interaction of two communities with a different history, traditions,
languages and in great measure, objectives.

\subsubsection{Numerical Relativity and Relativistic Astrophysics.}
\label{s:NRvsRA}
As somehow hinted in the previous paragraphs, it is natural to assume
that a ``generic'' relativist would have interest in both studying the
nature of solutions of theory ---and hence improving his or her
understanding of the theory--- and also of making contact with the
observation ---which then would help to set the context and validity
of the theory. Given the broad spectrum of research topics that
relativists deal with, it is to be expected that the particular choice
of research problem will set the emphasis on either conceptual or
observational aspects. In particular, this applies to the case of a
numerical relativist. For example, from a simplified point of view a
``gravitational collapse code'', can be potentially read as a tool to
either understand conceptual issues like Cosmic Censorship or as a
tool to predict astrophysical phenomena such as supernovae explosions.
In the first case, it is natural for the numerical relativist to gain
insight and even try to simplify the code through a geometric view
point. In the second case, she or he has no option but to attempt to
include all physical ingredients that are believed to play a role.
Therefore, geometric insights may be comparatively less relevant. In
any case, whether a geometry point of view is used or not reflects,
ultimately, in the way the code is constructed and read. Summarising, a given
problem in numerical relativity can be labelled as either
\emph{geometric} or as \emph{astrophysical} numerical relativity. The
tenet of this review is that the effects of the interaction between
mathematical and numerical relativists is bound to be more fruitful in
the former case, while in the latter the rewards would be more
indirect. This duality will be a recurrent theme in the review.

\subsection{Tensions and synergy between Geometry and Numerics}
As a first contact with the topic of the review we present some
general reflexions which illustrate the possible tensions and
synergies to be expected as result of the interaction between geometry
and numerics.

\paragraph{Global versus local.}
Global issues dominate modern mathematics. On the other hand, most
numerical simulations are local in time. This leads to a tension
between the tools and goals in each community.  An example of how
different the points of view could be is given by the notion of black
hole. The classical definition of a black hole makes use of global
ingredients ---e.g. \cite{Chr03}.  On the contrary, \emph{everyday}
numerical simulations make use of quasi-local characterisations related
to the notion of trapped region. This tension reaches its climax in
statements like \emph{there are no results about the existence of
  asymptotically flat vacuum black hole spacetimes with radiation}
---see the discussion in \ref{s:BHs}--- by a mathematical relativist
that can provoke the smile of a numerical relativist. However, this
tension is not as dramatic as it appears at first sight. On the one
hand, it is very possible that numerical simulations of binary black
holes could provide the missing clues for an eventual global existence
proof of the existence of dynamical black hole spacetimes
\cite{Pre07}. On the other hand, the global framework provided by the
Cosmic Censorship Conjecture acts precisely as guarantor of some of
the technical tools used in numerical simulations ---for example
excision or punctures.

\paragraph{Good geometric or analytic properties as a hope for good numerics.}
To a geometer who is inexperienced in the field of numerics, one of
the first reality checks he or she has to confront is that analytic
well-posedness and convergence of a numerical solution to the analytic
solution, do not guarantee the long term stability of the simulation
---this issue has been clearly discussed in \cite{BabSziWin06}. This
last point raised before the big break-throughs in the simulation of
black hole binaries is still valid as regards to the realistic and
accurate extraction of physics \cite{Pre05,CamLouMar06,BakCenCho06}.
For this, it is possible that an analytic or geometry insight may
provide the crucial ingredient.

\paragraph{Geometry as a way of prescribing and extracting physics.}
Numerics relies on the use of coordinates. This results, unavoidably,
in ambiguities in the extraction of physics if geometric notions are
not employed. However, these methods are global and hard to implement
numerically.  In practical applications, one could use, for example,
perturbative or post-Newtonian notions to extract the physics.
Nevertheless, this approach already relies in the acceptance of a
certain global behaviour of the spacetime. But more importantly, the
recent advances in the numerical simulations of spacetimes could
provide the opportunity for the reassessment of geometrical objects as
tools for the extraction of physics.

\medskip
We conclude the section presenting some examples which 
give a taste of both the difficulties and rewards of the
interaction between geometry and numerics:
\begin{itemize}
\item[-] \emph{Application of ideas from dynamical horizons to the extraction
    of physical parameters.} The development of some quasi-local approaches to black horizons
    ---see section \ref{s:quasi_bh}--- has been directly inspired by the
    needs of the numerical implementations. In particular Dynamical Horizons
    have already provided first examples of successful interaction in the
    calculation of mass, angular momentum and associated parameters
    ---cf. Krishnan's IHP talk and
\cite{Bai_al04,SchKri05,SchKriBey06,KriLouZlo07,CamLouZlo07}.      
  \item[-] \emph{Semiglobal evolution from small data.} The existence
    results of the development of hyperboloidal initial data close to
    Minkowski by Friedrich \cite{Fri86b} has been illustrated by the numerical
    simulations of H\"ubner ---see \cite{Hub01b}.
\item[-] \emph{Hawking mass.} The straight-forward implementation of some
  natural geometric notions meets often with numerical difficulties. This has
  been exemplified for the case of the Hawking mass in Schnetter's IHP talk \cite{IHP}.
\item[-] \emph{Calculation of high order derivatives.} Any numerical
  simulation contains noise which tend to be amplified in the
  numerical evaluation of derivatives ---see again Schnetter's IHP
  talk \cite{IHP}. This, for instance, complicates the implementation
  of constructions involving derivatives of the curvature tensor
  ---like in the case of the metric equivalence problem in
  GR \cite{Kar80a,Kar80b}. 
\item[-] \emph{Handling of coordinates.} Sometimes the use of coordinates
  which are easy to handle from a numerical point of view, could lead
  to complications like coordinate singularities and/or singular
  equations ---see e.g. Schnetter's IHP talk \cite{IHP}. In some
  circumstances a detailed geometric analysis does suggest a
  reformulation of the problem which would solve this particular
  tension \cite{Ans05,Ans06}.
\item[-] \emph{Penrose inequality.} Classical bounds on energy loss in
  the head-on collision of black holes using the geometric Penrose
  inequality ---see e.g. \cite{Gib72}--- have been shown to be overtly
  optimistic by numerical simulations. On the other hand, lines of
  thoughts based on the Penrose inequality have been useful in the
  discussion of initial data \cite{DaiLouTak02}.
\item[-] \emph{Geometry-inspired initial data sets.} Attractive geometric
  properties have been employed and prescribed in the construction of
  initial data sets for black hole spacetimes ---see among others e.g.
  \cite{Dai01a,Dai01c,Reu88,JezKijLes07}. However, the numerical
  usefulness of these geometric inputs is still to be assessed.
\item[-] \emph{Geometry-inspired evolution systems.}  The close
  interaction between geometry and numerics is illustrated in the
  application of geometry as a guideline in the development and
  implementation of certain GR evolution systems ---e.g. Z4 formalism
  in \cite{BonLedPal03} and also \cite{Rin06,RinSte05,Zen06,Zen07}.
  However, the role of Geometry is not so evident in most numerical
  codes ---see section \ref{s:evol_formalism}.
\end{itemize}

It may be of interest for the reader to look at the list of
successes given in \cite{And06}.

\subsection{Cautionary notes}
Although this review stems from a workshop with a similar name \cite{IHP}, it
should not be regarded as a proceedings or a systematic account of the
former. We have used the diverse contributions to inspire a reflection
on the interaction between geometric or analytic methods and numerical
tools in the context of GR. In this sense, unavoidably, it projects
our personal prejudices and it is certainly not comprehensive. We are
indebted to all the participants for the effort put on their
presentations and apologise beforehand in case of any
misrepresentation ---or omission--- of their ideas. The presentations
during the workshop will be referenced in a bundle as \emph{IHP
  talks}.

It is clear ---although it is already evident from the introductory
paragraphs--- that the word \emph{Geometry} has been used in a loose sense that
encompasses more than Differential Geometry and includes, for example,
(global and local) analysis, PDE theory or Group theory.  

As already mentioned, our discussion will be centred in numerical and
mathematical aspects of isolated systems in GR. There are other
streams of Relativity which have experienced a fruitful interaction
between the analytical and numerical communities: namely, on the one
hand higher dimensional spacetimes and on the other the study of
Cosmological spacetimes. In the latter case, this interaction has been
reported in the review \cite{Ber02} ---see also the discussion in
\cite{And06}.  Without going into much detail, we just mention that
the analysts (geometers) working in mathematical Cosmology have gained
much from insights obtained through numerical experiments in which,
for example, the behaviour of Gowdy and Bianchi IX spacetimes is
explored towards the initial singularity ---for a recent work on the
subject see \cite{Bey07}. In particular, it was using this sort of
numerical experiments that the existence of the so-called spikes
---non-smooth behaviour and big gradients of certain field quantities
as one approaches to the singularity--- was firstly attested in the
Gowdy spacetimes \cite{Ber02}. The analytical existence and nature of
these spikes has since been thoroughly analysed. This is a prime
example of the use of numeric techniques as a tool to explore the
nonlinear behaviour of the gravitational field.

We start from the premise that Mathematical Relativity and Numerical
Relativity are research areas on their own right. This review is addressed to
those researchers working in any of these two communities who feel that 
there is place for a fruitful interaction. We assume a broad general knowledge
of GR. We will not be extremely rigorous in our mathematical presentation and
will refer the reader to the specialised literature.

The article is structured as follows. Section \ref{s:IsolSysGR}
reviews some conceptual issues concerning the notion of isolated
bodies in GR, including asymptotic flatness and the black holes.
In section \ref{s:IVP} the initial value problem in GR is presented as
the appropriate approach to the generic construction of spacetimes. In
particular, \ref{s:diff_IVP} reviews the different types of initial
value problem.  Section \ref{s:initial_data} explores some aspects of
the construction of solutions to the constraint equations which we
believe are, or could be, of interests for numerical applications
---the topology of initial hypersurfaces, the conformal method, the
thin sandwich and the recent developments in gluing constructions.
Section \ref{s:evol_formalism} discusses various aspects of evolutions
formalisms. Here, particular emphasis is given to initial value
problems of Cauchy type, although some observations concerning the
characteristic and hyperboloidal problems are made. There is, also, a
discussion of the conformal field equations. Section \ref{s:art_bh}
reviews some aspects of the modern description of black holes
including recent results about uniqueness and rigidity of the Kerr
spacetime, quasi-local definitions of black holes, geometric
inequalities, helical Killing vectors and aspects of ``puncture''
black hole evolutions. Section \ref{s:miscellaneous} discusses
miscellaneous topics that did not find an adequate place in the
structure of the review, but that we believe could not go unmentioned.
Finally, \ref{s:conclusions} provides some conclusions and a brief
list of topics where, we believe, the interaction between Geometry and
Numerics will acquire particular relevance in the coming years.

\section{Isolated systems in General Relativity}
\label{s:IsolSysGR}

As mentioned before, this review will be concerned with solutions of General Relativity 
describing isolated systems. GR is formulated as a
gauge theory of gravity, usually in terms of the
\emph{Einstein field equations} (EFE) 
\bea
 \label{e:EFE}
 R_{\mu\nu}-\frac{1}{2}Rg_{\mu\nu}= 8\pi T_{\mu\nu} \ \ , 
\eea 
where
 $g_{\mu\nu}$ denotes the metric of the spacetime, $R_{\mu\nu}$ its
 corresponding Ricci tensor, $R$ the Ricci scalar and $T_{\mu\nu}$ is
 the so-called stress-energy tensor ---in geometric units $c=G=1$.
 Solutions to the EFE correspond to geometries rather than to metrics
 given in a specific coordinate system, since metrics related by
 spacetime diffeomorphisms are physically equivalent. As a consequence
 of this, equation (\ref{e:EFE}) understood as a  partial differential 
equations (PDE) system for the
 metric components $g_{\mu\nu}$ constitutes an underdetermined
 (incomplete) system that cannot produce a unique solution.  An extra
 ingredient, namely a \emph{gauge fixing} procedure, must be provided
 to establish a standard PDE problem ---note this remark applies to any 
geometrically introduced PDE.
 Here we adopt a broad notion of gauge which includes not only a
 coordinate choice but could also contemplate, for example, a tetrad
 choice and even perhaps a choice of conformal factor.

We are interested in solving equation (\ref{e:EFE}) subject to
boundary conditions describing an isolated system for which the
effects of the Cosmological expansion are neglected. A first way of
approaching this problem is the construction of exact solutions to
equation (\ref{e:EFE}) by making assumptions on the symmetries of the
spacetime and/or on the algebraic structure of their curvature tensors
---the most comprehensive treatise on exact solutions is \cite{SKMHH};
reviews concerned with the physical aspects of exact solutions can be
found in \cite{Bic00,Bic06}.  This way of doing things has provided
deep conceptual and physical insight, as it has supplied the tools of
most of the observational predictions of GR. Furthermore, it is the
basis of perturbative analysis.

However, if one is interested in making statements about generic
spacetimes having the right kind of boundary conditions and on how to
construct them in a systematic manner, then one is led to consider an
\emph{initial value problem} or more generally and \emph{initial
  boundary value problem}. Initial value problems can be of
\emph{Cauchy} type ---that is, the initial data is prescribed on a
\emph{Cauchy hypersurface}\footnote{For a discussion of the
  relationship between global hyperbolic spacetimes and the existence
  and properties of smooth Cauchy surfaces, in particular the
  existence of Cauchy time functions, see
  \cite{BerSan03,BerSan04,BerSan06}.}  --- or not ---like in the case
of the \emph{characteristic} and the \emph{hyperboloidal} \cite{Fri83}
initial value problems ---for a more complete discussion on these
ideas see subsection \ref{s:diff_IVP}.  Only Cauchy initial value
problems allow, in principle, to reconstruct the whole spacetime,
while the characteristic and hyperboloidal problems allow at most to
reconstruct the domain of influence of the initial hypersurface.

The overall strategy is to prescribe suitable initial data on the
initial hypersurface and then to evolve them by means of the EFE and
the appropriate equations for the matter part. This is very much in
the spirit of the approach taken by numerical relativists. There are,
however, a number of caveats to this approach ---like existence and
uniqueness of solutions, stability and global issues including Cosmic
Censorship--- which are the concern of mathematical relativists (see
\cite{Sanch06} for a review addressing some open problems in
mathematical relativity and articulated around the notion of Cauchy
hypersurface). In particular, the satisfactory resolution of these
caveats by the mathematical community is the guarantor of the
consistency of the strategy employed by the numerical relativists.

Very often, the needs of the numerical community are well ahead of the
developments of the mathematical community.  This applies, in
particular, to the study of the motion of isolated bodies in GR, a
problem of clear interest for astrophysicists which has been dealt
numerically for a long time now ---see e.g. \cite{BauSha02}.  The
existence issue, that is taken for granted in the numerical community,
is a challenging mathematical problem. We emphasise that this last
point represents a crucial aspect in the overall coherence of the
problem. The afore-discussed state of affairs is exemplified by the
recent work on the existence of solutions describing the motion of
isolated bodies ---dust balls in this case--- by Y. Choquet-Bruhat and
H.  Friedrich \cite{ChoFri06}. See also Choquet-Bruhat's IHP talk
\cite{IHP}. Finally it must be said that a converse phenomenon is also
true. The numerics for many mathematical questions is still
awaiting to be developed ---for example the well-posed, constraint
preserving initial boundary value problem constructed by Friedrich \&
Nagy \cite{FriNag99}.

\subsection{Asymptotic flatness}
\label{s:asymflat}

The physical intuition suggests that the spacetime describing an
isolated body far away from it, should be Minkowskian in some
appropriated sense. Such a behaviour is generically referred to as
\emph{asymptotic flatness}. As already mentioned, checks on the
consistency of the theory requires a rigorous formulation of this
physical intuition.

The metric of an asymptotically flat spacetime, when expanded in terms
of some convenient distance parameter, should coincide with the flat
Minkowski metric at leading order. The physics of the problem is, on
the other hand, encoded in the low order terms. In particular, one
should expect to find there information about the mass and radiative
properties of the system. Now, it is natural to ask, firstly, whether
it is possible to rephrase the above statements in a coordinate
independent way; and secondly, if one can make generic statements
about the properties of the low terms in the asymptotic expansions
of the metric of an isolated system. To this end, Penrose introduced
the more specific notion of \emph{asymptotic simplicity} ---see e.g.
\cite{Pen63,PenRin86} for a precise definition. Roughly speaking, a
spacetime $(\mathcal{M},g_{\mu\nu})$ is said to be asymptotically
simple if there exists a positive scalar $\Omega$ (the conformal
factor) and a spacetime with boundary
$(\hat{\mathcal{M}},\hat{g}_{\mu\nu})$ ---the compactified, unphysical
spacetime--- with $\hat{g}_{\mu\nu}=\Omega^2 g_{\mu\nu}$ such that the
boundary of $\hat{\mathcal{M}}$, defined by $\Omega=0$, has a structure
similar to that of the standard compactification of Minkowski
spacetime ---that is, the conformal boundary consists, at least, of a
null hypersurface (null infinity) $\scri=\scri^-\cup \scri^+$. In the
particular case of the Minkowski spacetime, the conformal boundary
consists, in addition, of three points: $i^0$ ---spatial infinity---
and $i^-$ and $i^+$ ---respectively, past and future timelike
infinity. In general, these additional points will not be present in
the conformal completion of a generic asymptotically simple spacetime.
However, if the spacetime is constructed as the development of 
asymptotically Euclidean initial data ---see equation
(\ref{asympt_Euclidean})--- then $i^0$ will be part of the
conformal boundary by construction. The situation with respect to
$i^-$ and $i^+$ is more complicated. If the relevant spacetimes
contain black holes, then there is no reason to expect that $i^\pm$ be
present in the conformal completion, or if they are, that their
structure should be that of a point ---this is a difficult open
question which will benefit from input from numerical simulations. In
any case, and to summarise, the definition of asymptotic simplicity
gets around these issues by being only concerned with null infinity.
The definition of asymptotic simplicity is geometric in what it avoids
the use of coordinates.  Penrose's original definition requires the
boundary of the unphysical spacetime to be smooth ($C^\infty$).
Penrose's original idea was to use the notion of asymptotic simplicity
as a way of providing a characterisation of isolated systems in GR:
\emph{ the far fields of spacetimes describing isolated systems should
  be asymptotically simple in the sense that they admit a smooth
  conformal compactification to null infinity}. The latter suggestion
is known in the literature by the name of \emph{Penrose's proposal}.

The conceptual and practical advantages of the use of the compactified
picture to describe isolated systems have been discussed at length in
the literature ---see e.g.
\cite{Ash84a,Fri92,Fri98b,Fri99,Fri03a,Ger76}. Here we just mention
that: i) the use of the compactified picture allows to rephrase
questions concerning the asymptotic decay of fields by questions of
their differentiability at the conformal boundary; ii) if the
spacetime is asymptotically simple, then it is possible to deduce a
certain asymptotic behaviour for the components of the Weyl tensor
known as the \emph{peeling behaviour} ---see e.g. \cite{Fra04}.  If
the spacetime is such that it would not admit a smooth
compactification ---say, just $C^k$ instead of being $C^\infty$--- then
part of this formalism can still be recovered, but one has to be more
careful ---see e.g.  \cite{ChrMacSin95}.

The idea of asymptotic flatness and the notion of asymptotic
simplicity are inspired by the analysis of exact solutions to the EFE.
If one wants to establish more general properties of this class of
spacetimes one has to resort to a formulation of the problem based on
an initial value problem and then make use of analytic techniques to
obtain qualitative information about the solution or construct the
solution numerically. Here, again, intuition suggests that the right
initial conditions for obtaining an asymptotically flat spacetime are
\emph{asymptotically Euclidean} ones. More precisely, one could
require an initial hypersurface $\Sigma$ with at least one
asymptotic end (a region which is topologically $\Real^3$ minus a ball)
in which one can introduce coordinates $x^i$ such that 
\begin{equation} \label{asympt_Euclidean}
\gamma_{ij}=\left(1 + \frac{2m}{|x|}\right)\delta_{ij} +\mathcal{O}\left(\frac{1}{|x|^2}\right), \quad K_{ij}=\mathcal{O}\left(\frac{1}{|x|^2}\right),
\end{equation}
where $|x|=\sqrt{(x^1)^2+(x^2)^2+(x^3)^2}$, and $\gamma_{ij}$ and $K_{ij}$
are, respectively, the first and second fundamental forms of
$\Sigma$ ---see section \ref{s:initial_data} for more on this.
More general notions of asymptotic Euclideanity have been considered
in the literature ---see e.g.  \cite{Cha82,ChaCho80,ChrOMu81}. The one
given here, has a well defined ADM mass at the asymptotic end in
question.

The first results concerning the existence of solutions to the vacuum
EFE under the above boundary conditions can be traced back to
\cite{Cho52} in which local existence in time was shown to hold
---well-posedness.  Choquet-Bruhat and Geroch introduced the notion of
a maximal future development of an initial data set, and showed that
for each given initial data set there is a unique maximal future
development \cite{ChoGer69}.  The work by Penrose has given a negative
answer to the question of completeness in his singularity
theorem \cite{Pen65b}: given initial data where the initial Cauchy
hypersurface is non-compact and complete, if the hypersurface contains
a closed trapped surface ---see section \ref{s:trapped_surfaces}---
 then the corresponding
maximal future development is geodesically incomplete. A further
milestone was the resolution of the so-called \emph{boost problem in
  GR} given in \cite{ChoChr81,ChrOMu81}. Along these lines, the
existence results of Friedrich \cite{Fri86b} where it has been shown that
the development of initial sufficiently small data given on an
hyperboloid is complete. It should be noted that a hyperboloid is not
a Cauchy hypersurface, and in this sense the results of \cite{Fri86b}
are \emph{semiglobal}. The analysis of the non-linear stability of the
Minkowski spacetime of \cite{ChrKla93} has shown, among other things,
the existence of vacuum radiative spacetimes which are asymptotically
flat. More precisely, they show that every asymptotically flat initial
data which is globally close to the trivial data gives rise to a
solution which is a complete spacetime tending to the Minkoswki
spacetime along any geodesic. There are no additional restrictions on
the data. The result gives a precise description of the asymptotic
behaviour at null infinity. Remarkably, however, the analysis is
insufficient to guarantee the peeling behaviour of the solutions. An
alternative proof has been provided by Lindblad \& Rodnianski
\cite{LinRod04}. This provides a global stability result of Minkowski
spacetime for the vacuum EFE in wave coordinate gauge ---see section
\ref{s:GH}. The initial data coincides  with the Schwarzschild solution
in the neighbourhood of spacelike infinity.  This result is less
precise as far as the asymptotic behaviour is concerned. More recently,
following the spirit of Christodoulou \& Klainerman's proof, Bieri
---see her IHP talk \cite{IHP}--- has considered more general
asymptotically Euclidean initial data ---i.e.  with less decay and one 
less derivative--- and shown the existence of a solution which is a
geodesically complete spacetime, tending to the Minkowski spacetime at infinity
along any geodesic. The solutions have finite ADM energy, but the
angular momentum on the initial hypersurface is not necessarily well
defined. The question answered by Bieri's analysis was what is the
most general non-trivial asymptotically flat initial data set giving
rise to a maximal development that is complete. The remaining open
question in this programme is to find sharp criteria for non-trivial
asymptotically flat initial data sets to give rise to complete
maximal developments.

In what concerns the existence of spacetimes which satisfy the peeling
behaviour, Klainerman \& Nicol\`o \cite{KlaNic03a,KlaNic03b} have
shown that the peeling behaviour is satisfied by the development of a
big class of asymptotically Euclidean spacetimes. This result is,
however, not sharp for it can be seen that initial data for the Kerr
spacetime ---which satisfies the peeling behaviour as it is a
stationary spacetime, see e.g. \cite{DamSch90,Dai01b}--- is not
contained in the class of initial data sets they considered.

Regarding the related question of the existence of asymptotically
simple spacetimes, it is now known that there is indeed a big class of
spacetimes with this property ---see \cite{ChrDel02}.  The essential
ingredients for the latter result are the semiglobal results of
\cite{Fri86b}, and a refined version of the initial data sets which
can be constructed using the Corvino-Schoen method.  This construction
allows to glue fairly arbitrary asymptotically Euclidean initial data
sets to a Schwarzschild asymptotic end ---for more on this gluing
construction see section \ref{s:corvino_gluing}. It has to be mentioned that
the class of spacetimes obtained in \cite{ChrDel02} are very special
because by construction, they do not contain any radiation near
spatial infinity. The fact that the initial data is Schwarzschildean
near infinity, implies that the Newman-Penrose constants of the
spacetime vanish ---the Newman-Penrose constants are absolutely
conserved quantities defined in the cuts on null infinity
\cite{NewPen65,NewPen68}. This value propagates along the generators
of null infinity, and it allows us to conclude that the Weyl tensor
vanishes at future timelike infinity $i^+$ ---see
\cite{FriSch87,FriKan00}.

Given the state of affairs described in the last paragraph, the challenge
is now to find the sharp criteria on an initial data set to obtain an
asymptotically simple spacetime. Seminal work in this direction is
\cite{Fri98a} ---see also \cite{Fri04}--- where a convenient framework
to discuss the behaviour of the region of spacetime where null and
spatial infinity meet ---the so-called \emph{cylinder at spatial
  infinity}--- has been introduced. It is important to point out that
this work and related ones ---see e.g.
\cite{Val04a,Val04d,Val04e,Val05a}--- are carried out in the conformally
rescaled spacetime (unphysical spacetime) and make use of the
so-called \emph{conformal field equations}, a generalisation of the
EFE which exploits the extra gauge freedom ---a conformal one--- that
has been introduced via the notion of asymptotic simplicity. The
programme started in \cite{Fri98a} and continued in
\cite{Fri04,Val04a,Val04d,Val04e,Val05a} has showed that there is a
hierarchy of \emph{obstructions to the smoothness of infinity} which
control the smoothness of the spacetimes at null infinity ---see also
Valiente Kroon's IHP talk \cite{IHP}. These obstructions are
expressible in terms of parts of the initial data of the spacetime. A
further analysis of the obstructions points towards the following
conjecture: \emph{ if the development of asymptotically Euclidean
  initial data admits a smooth conformal compactification (in the
  sense of Penrose) at both past and future null infinity, then the
  initial set is stationary near infinity} ---see
\cite{Val04a,Val04d,Val04e,Val05a}. If shown to be true, the latter
conjecture would give a prominent role to the initial data sets which
can be constructed by means of the Corvino-Schoen gluing construction.
It is worth mentioning that the developments of conformally flat
initial data sets like the Brill-Lindquist \cite{BriLin63}, Misner
\cite{Mis63}, Bowen-York \cite{BowYor80} and Brandt-Br\"ugmann
\cite{BraBru97}, as well as the data used systematically in the
numerical simulations of binary black hole mergers ---see the end of
section \ref{s:helical}--- will have a null infinity (if any) which is
not smooth. However the differentiability of the conformal boundary
seems to be enough to guarantee the peeling behaviour \cite{Val07a}.

Given an arbitrary asymptotically Euclidean initial data set $(\Sigma,\gamma_{ij},K_{ij})$,
using the Corvino-Schoen construction one could construct another
initial data set $(\Sigma',\gamma_{ij}',K_{ij}')$ which coincides with
the former inside a compact set, but which is exactly stationary near
infinity.  The development of $(\Sigma,h_{ij},K_{ij})$ will possess a
non-smooth but possibly peeling null infinity. On the other hand, the
development of $(\Sigma',h_{ij}',K_{ij}')$ will be smooth. A natural
question is the following: is there any substantial difference between
the physics of the two developments? This is a complicated question
which will only be answered in a satisfactory manner by a close
collaboration between numerical and mathematical relativists. It is
very likely that the latter issue will not have any relevance for
numerical relativists interested in astrophysical applications. On the
other hand, they provide a whole unexplored area for those who want to
use the numerical methods to understand the nature of GR.

\subsection{Black holes} 
\label{s:BHs}
A black hole is a region of no escape which does not extend out to
infinity.  Traditionally ---see e.g. \cite{HawEll73}--- the definition
is made using the conformal compactification of spacetime. Namely, if
the spacetime admits a suitable null infinity and the causal past of
null infinity is globally hyperbolic ---i.e. it admits a Cauchy
hypersurface--- then one defines the black hole region like the
spacetime minus the causal past of future null infinity. The boundary
of the black hole is the event horizon. A spacetime containing a
single black hole contains two natural boundaries, one is null
infinity and the other is the event horizon ---an inner boundary. The
original picture arose from an analysis of spherical symmetry.
According to the so-called weak Cosmic Censorship picture ---cf.
section \ref{s:geom_ineq}--- the end state of any generic
asymptotically Euclidean initial data set will be a particular
stationary black hole spacetime: the Kerr solution. Naked
singularities can actually form, but the resulting spacetime are
unstable \cite{Chopt92,Chr94,Chr99} and non-generic. The rigorous
definition of a black hole makes use of global ingredients: in
particular it requires the existence of a complete $\scri^+$. As
already mentioned, there are no global existence proofs of dynamical
black hole vacuum spacetimes. The latter is a point deserving further
clarification.  There are results concerning the global existence of
dynamical black holes in, for example, the spherically symmetric
Einstein-scalar field theory ---e.g.  \cite{Daf03} for the formation
of black holes from a regular past--- or in the vacuum 5-dimensional
setting. Regarding the latter example, see \cite{DafHol06} where the
\emph{orbital} non-linear stability ---i.e. with respect to
perturbations respecting symmetry--- of the Schwarzschild-Tangherini
black hole is discussed, and thus the existence of a global,
non-stationary black hole is established.  Unfortunately, these
examples are not relevant for astrophysics. The available
\emph{examples} of 4-dimensional vacuum dynamical black holes suffer
from some \emph{pathology} ---like the Robinson-Trautman spacetimes
discussed in \cite{Chr92a} with a complete future null infinity, but
where past null infinity is incomplete--- or are local ---like in the
discussion of ``multi-black hole'' spacetimes of \cite{ChrMaz03}, or
that of spacetimes with isolated horizons of \cite{Lew00}.

Among other things, a proof of the
nonlinear stability of the Kerr spacetime would show the
existence of a wide class of spacetimes with non-stationary black
holes. The idea behind such a proof is to show that initial data which
are in some sense close to Kerr initial data will give rise to a black
hole spacetime whose global structure will be similar to that of the
Kerr spacetime. In addition to proving the existence of dynamical
black hole spacetimes, such a result would be an important step
towards a proof of the (strong) Cosmic Censorship Conjecture ---see
section \ref{s:IVP}. A particular situation in which the latter
conjecture has been proved in an asymptotically flat context is the
spherically symmetric Einstein-scalar field system studied by
Christodoulou ---see \cite{Chr86,Chr87,Chr91,Chr92,Chr99}. A crucial
property of stationary black holes is the so-called \emph{rigidity}:
that is, the fact that stationarity together with a couple of further
assumptions on the behaviour of the horizon imply further symmetries
in the spacetime ---i.e. axial symmetry. The ideas behind the rigidity
theorems have been recently revisited by Isenberg \& Moncrief with the
aim of obtaining generalisations which are valid for higher dimensions
---cf. Isenberg's IHP talk \cite{IHP} and section \ref{s:rigidity}.

\subsection{Critique and practicalities} \label{s:critique}

There has been a number of critiques to the notion of asymptotic
flatness, most notoriously the ones articulated by Ellis
\cite{Ell84,Ell02}. This critique has Cosmological motivations and
arose from the desire of analysing the nature of the interaction
between local physics and boundary conditions in the expanding
Universe. In particular, there was a desire to understand what
difference does the use of particular boundary conditions make on
results depending in an essential manner on asymptotic flatness,
like the peeling behaviour, the positivity of mass ---cf. section
\ref{s:geom_ineq}--- and black hole uniqueness. Ellis argues, for
example, that it is unrealistic when considering local gravitational
collapse to worry about observations with infinite life-times. The
latter is what is implied in the usual definition of a black hole
---see however section \ref{s:quasi_bh}. The alternative to null
infinity as a natural boundary of spacetime given in \cite{Ell84} is
to consider a world-tube ---\emph{finite infinity}--- with a timelike
boundary located at a spatial radius which is sufficiently far away
from the sources. If brought to the context of an initial value
problem, Ellis proposal implies the use of an initial boundary value
problem to describe the physics of isolated systems in which the
effects of the Universe are put in by appropriate boundary conditions.
The well possedness of the initial boundary value problem has been
firstly discussed in \cite{FriNag99}. The analysis given there shows
that if one makes use of a finite infinity then there is no covariant
way of prescribing the boundary information. This, in turn, would lead
to ambiguities in the extraction of physics.  One of the rationales
behind the introduction of the \emph{null infinity formalism} was
precisely to avoid this sort of problems.  For a neat discussion of
these issues, with an emphasis to the extraction of radiation see
\cite{Fra00}.

This critique is remarkable in what that if one takes it away from its
Cosmological context it could also be employed when addressing the
relevance of global issues for numerical relativity. Most numerical
simulations make an implicit use of the notion of finite infinity by
calculating numerically the solution to an initial boundary problem.
In this approach there is always the potential of ``messing up'' the
physics of the problem by setting boundary conditions which are not
appropriate ---see section \ref{s:outer_boundaries}. There are,
however, also simulations which compactify in space, see \cite{Pre07}.
Moreover, the numerical simulations do not cover an infinite time
although given the current advances in the field there seems to be no
essential difficulty for letting the simulations ``to run as long as
it is necessary''.

It is worth noticing that there have been some semiglobal numerical
calculations using the conformal field equations which exemplify the
hyperboloidal existence results of \cite{Fri86a} ---see
\cite{Hub99a,Hub99b,Hub01a,Hub01b}. There are some further attempts to
calculate portions of the Schwarzschild from hyperboloidal data
---\cite{HubWea01} unpublished. For a critical review of the successes
and problems of this programme see \cite{Hus03}. More recently, there
have been some global calculations of the Schwarzschild spacetime
\cite{Zen06,Zen07}.  These calculations make use of a more general
version of the conformal field equations and conformal Gauss
coordinates. These numerical simulations exemplify the analysis of
this type of coordinates given in \cite{Fri03c}. We also mention the
recent fully pseudo-spectral scheme \cite{HenAns08} for solving hyperbolic
equations on conformally compactified spacetimes.

The discussion in the previous paragraphs leads to a situation where
besides a number of simulations tailor-made for exemplifying some
analytical aspects of solutions of the EFE, a majority of the
state-of-the-art numerical codes used to simulate spacetimes
containing dynamical black holes are not designed for addressing
global issues in GR. So, what is the relevance for the numerical
community of the work on global issues carried out by mathematical
relativists? As mentioned in the introductory section, the role is to
ensure that some of the approaches implemented in numerical
simulations are justified. For example, if the non-linear stability of
the Schwarzschild turned out to be false, then it could happen that
the development of an asymptotically Euclidean black hole initial
data would not have the asymptotic structure of the Kerr spacetime,
and thus measurements of gravitational radiation at finite radius
would be rethought. Similarly, if the no-hair theorems were not valid,
there would be no justification for employing perturbative methods on
a Kerr background to analyse the late stages of the evolution of a
black hole binary merger. These examples are extreme but highlight the
need of having the theoretical framework on a sound ground. 

Finally, we would like to retake the point raised in \cite{And06} that
the problem of the stability of the Kerr spacetime opens a natural
arena for the interaction of numerical and mathematical relativity.
Indeed, it is easy to imagine that tailor-made simulations could
provide crucial insights concerning the asymptotic decay of the
gravitational field near the horizon. This interaction would be,
notwithstanding, on the lines of \emph{from Numerics to Geometry}.

\section{Initial value problems} \label{s:IVP}

If one wants to discuss generic properties of solutions to the
Einstein field equations which describe in some appropriate sense
isolated systems, one has to find a \emph{systematic} way of obtaining
these solutions. The only approach to the construction of solutions
which seems general enough is that of an initial value problem (IVP).
This mathematically sound approach is perhaps not ideal from a
physical point of view, and indeed a number of objections have been
raised. One could argue that all major observational predictions of GR
have been obtained by means of the analysis of exact solutions.

A first (conceptual) objection consists in making sure that one is
able to reconstruct, starting from initial data, the whole maximal
extension of the spacetime ---the strong Cosmic Censorship
\cite{Pen79,Pen82b,And04}. This has implications on causality
violations and predictability of the theory ---see e.g. \cite{Sanch06}.

Another objection questions to what extent an approach to GR based on
an initial value problem is appropriate to obtain predictions which can
be contrasted with measurements. One of the crucial problems behind
is that for a certain system of physical interest, there may be many
possible ways of constructing \emph{physically plausible} initial
data. This is particularly the case when attempting to construct
initial data for spacetimes containing \emph{dynamical} black holes.
In order to construct data for a black hole spacetime, one should be
able to find a way of identifying the black hole in the initial data.
For these purposes, the notion of event horizon is of no use for it is
a global quantity which is only known once the whole spacetime ---up
to null infinity--- is known. Instead, one can use the notion of
apparent horizon: for if there is an apparent horizon on the data it
must be contained in the black hole region ---see section
\ref{s:quasi_bh}. In other words, the presence of an apparent horizon
indicates the existence of a black hole ---if the weak Cosmic
Censorship is true.  Furthermore, data with an apparent horizon will
be geodesically incomplete as a consequence of Penrose's singularity
theorem \cite{Pen65b} ---thus, indicating the formation of a
singularity. It should be emphasised that a black hole data does not
need to contain an apparent horizon, but if it contains one, it must
be a black hole. Similarly, it is not clear whether data with two
apparent horizons will contain two black holes. However, this is a
reasonable assumption to begin with. To make things worse, note that
many of the approaches for constructing initial data for black holes
are explained more in terms of mathematical advantages, like the
simplicity of the equations ---like in the case of conformal
flatness--- than in physically based considerations ---see however
section \ref{s:quasi_bh} for an account of the ideas introduced by the
quasi-local approach to black holes.  To summarise, physics seems to be
hard to encode in initial data sets for the EFE.

There has been a number of works devoted to compare the results of
numerical simulations in the search of robust aspects ---that
is, structures in the numerically calculated spacetimes which are
independent of the particular type of initial data being used ---see
e.g. \cite{BakCamPreZlo07,SpeBruGon07,ThoDiePol07} and IHP talks by Hannam
and O'Murchadha.  Crucially, the
usefulness of numerical simulations as a source of wave form templates
for the detection of gravitational radiation in the current generation
of interferometric gravitational wave observatories depends on the
assumption of the robustness. From the point of view of a mathematical
relativist or a geometer, this issue is essentially open and
unexplored. The problem is certainly complicated, and possibly one
will have to wait several years before some qualitative statement can
be made. The resolution of the non-linear stability of the Kerr
spacetime may provide first insights and rigorous answers to the
question of robustness of initial data sets.

\subsection{On the different types of initial value problems}
\label{s:diff_IVP}
Initial value problems can be classified according to the nature of
the hypersurface on which the initial data is prescribed. Usually, the
initial hypersurface $\Sigma$ is taken to be spacelike. As mentioned
in section \ref{s:asymflat}, when discussing initial value problems
for isolated systems the spacelike initial hypersurface $\Sigma$ is
naturally assumed to be asymptotically Euclidean. This particular type
of initial value problem in GR will be referred to as a \emph{Cauchy
  initial value problem} as it would allow, in principle, to recover
the whole spacetime ---see references \cite{HawEll73,Wal84} for a
rigorous discussion of these issues, and \cite{FriRen00} for a recent
review on the subject. The seminal work of Choquet-Bruhat \cite{Cho52}
has shown the well-posedness of the Cauchy initial value problem in
GR: an existence theorem local in time ---that is, existence is
guaranteed for a small time interval, in a neighbourhood of the
initial hypersurface. The portion of spacetime recovered by this local
existence theorem is called the \emph{development} of the initial
data. At this level is not possible to discuss global existence of
solutions without more assumptions or information on the initial data
---for further discussion on this, see for example the review
\cite{KlaNic99}.  Nevertheless, it has been shown that the development
has a unique maximal extension \cite{ChoGer69}.

Initial value problems on spacelike hypersurfaces are not necessarily
Cauchy initial problems. For example, the \emph{hyperboloidal initial
  value problem} first introduced by H. Friedrich in \cite{Fri83}, on
which initial data are prescribed on a spacelike hypersurface ---which
could be thought of as intersecting null infinity--- is not of Cauchy
type. An initial value problem on a hyperboloidal hypersurface would
not allow us to recover the whole spacetime. Moreover, it is not
possible to find out where the development of hyperboloidal data lies
in a globally hyperbolic spacetime, more specifically, one cannot relate 
the hyperboloidal data with a hypothetical Cauchy data. 

As pointed out in \cite{Stewa98}, the Cauchy problem in GR is in some
sense natural for theoretical discussions, but strictly speaking it
does not necessarily correspond to the way things are done in
numerical relativity, where one usually aims to solve the EFE only
within a compact domain ---cf. sections \ref{s:critique} and 
\ref{s:outer_boundaries}. The latter requires
the solution of an \emph{initial boundary value problem}, where in
addition to the data to be prescribed on a spacelike domain, one also
has to prescribe extra data on a timelike boundary. The presence of
this extra data ---the \emph{boundary data}--- raises the question of
which data can and must be specified on the timelike boundary in order
to have a solution ---and further, whether there are any compatibility
conditions required between the initial data and the boundary data. A
first rigorous discussion of the boundary value problem in GR, including a
local existence result has been provided by Friedrich \& Nagy
\cite{FriNag99}. A recent IVBP well-posedness result in the context of the
harmonic formulation ---see section \ref{s:GH}--- has been presented in \cite{KreWin06,KreReuSarWin07}.

The initial data must not necessarily be prescribed on spacelike
hypersurfaces. One could also prescribe it on null hypersurfaces, in
which case we speak of a \emph{characteristic initial value problem}.
However, as the domain of influence of a non-singular null
hypersurface ---i.e. a null hypersurface without caustics--- is empty,
one needs, for example, to consider data prescribed on two
intersecting null hypersurfaces in order to obtain a non-trivial
development. A local existence theorem for this sort of problem has
been given in \cite{Ren90}. A variant of the problem is to prescribe
data on a light cone. There is no existence result for the Einstein
equations in this case ---see however \cite{Fri86a}.

The characteristic initial value problem in GR can be traced back to
the seminal work of Bondi and collaborators on gravitational radiation
in the early 1960's ---see \cite{BonBurMet62,Sac62c}. An early systematic
discussion of the characteristic problem can be found in
\cite{Sac62b}. From the point of view of asymptotics, one can
formulate the so-called \emph{asymptotic characteristic initial value
  problem} in which initial data is prescribed on both an outgoing
light cone and on null infinity. The portion of spacetime to be
recovered from this initial problem lies at the past of these null
hypersurfaces. Existence results for this type of characteristic
initial value problem have been given in
\cite{Fri81a,Fri81b,Fri82,Kan96b} in the context of the conformal
field equations. An analysis of the well-posedness of the problem more
in the lines of the original work of Bondi and collaborators can be
found in \cite{FriLeh99}. Another type of characteristic initial value
problem is the $2+2$ initial value problem ---see e.g.
\cite{dInSma80}--- and the initial value problems with data prescribed
at past null infinity of \cite{Fri86a,Fri88}.

On a characteristic initial value problem, a part of the EFE
reduce to \emph{interior (transport) equations} on the null
hypersurface which can be recast as ordinary differential equations
along the generators of the null hypersurface. Further, the data are
free ---i.e. not subject to elliptic constraints like in the case of
spacelike hypersurfaces, see section \ref{s:initial_data}. This feature was
crucially exploited in the seminal work on gravitational radiation.
The characteristic problem is naturally adapted to the discussion of
gravitational radiation, however, it tends to run into problems when
caustics form. Further, physics is hard to prescribe on null
hypersurfaces.

Finally it is noted that an IVP of a mixed type on which data is
prescribed on a combination of spacelike and null hypersurfaces has
been considered: the \emph{Cauchy-characteristic} or \emph{matching
  problem} ---see \cite{Bisho93} and the review \cite{Win05}.

\subsection{Gauge reduction}
\label{s:gauge_red}
As it has been discussed at the beginning of section \ref{s:IsolSysGR}, the formulation
of the EFE as a standard PDE problem requires the adoption of a gauge
choice.  This {\em reduction process} ---see
e.g. \cite{Fri05,FriRen00}--- involves four different differential
systems:
\begin{itemize}
\item[i)] The \emph{main evolution} ---also, {\em reduced} or {\em relaxed}--- \emph{system}, whose
solution provides the spacetime geometry in a certain coordinate system. 
\item[ii)] The \emph{system of constraints}, which controls the permanence in
the submanifold of solutions of the theory.
\item[iii)] The \emph{gauge system}, which allows the fixation of a
  coordinate system and therefore casts the \emph{main evolution
    system} as a genuine PDE system.
\item[iv)] The \emph{subsidiary system}, consisting in the evolution
  equations of the auxiliary (subsidiary) constraints introduced in
  the previous steps and guaranteeing the overall consistency.
\end{itemize} 
The different manners of dealing with the previous points define a
variety of {\em evolution formalisms} in GR, where relative stress to
be placed on systems i)-iv) depends on the specific aspect of the EFE one
wants to address. In section \ref{s:evol_formalism} we will briefly
review some of the more relevant evolution formalisms considered in
numerical relativity with a focus on formulations based in the GR
Cauchy IVP. Prior to this, one must discuss the system of constraints
on a spacelike hypersurface.

\section{Initial data for the GR Cauchy problem}
\label{s:initial_data} 

The initial data to be prescribed in this IVP problem are given by a pair
of symmetric tensor fields $(\gamma_{ij}, K_{ij})$
on the initial $\Sigma$, respectively 
corresponding to the induced Riemannian metric and the extrinsic curvature
under the embedding of $\Sigma$ into  $(M,g_{\mu\nu})$.
Such an interpretation of $\Sigma$ as a spacelike hypersurface of 
$(M,g_{\mu\nu})$ demands the fulfillment of the so-called
\emph{Hamiltonian} and \emph{momentum} constraints, respectively
\begin{eqnarray}
&& {}^{(3)}\!R - K_{ij}K^{ij} + K^2 = 16 \pi \rho \label{Hamiltonian}\\
&& D_j\left(K^{ij} -\gamma^{ij}K\right) = 8\pi J^i  \label{Momentum},
\end{eqnarray}
where ${}^{(3)}\!R$ and $D_i$ are, respectively, the Ricci scalar and the
Levi-Civita connection associated with $\gamma_{ij}$, 
$K$ is the trace of $K_{ij}$, $\rho$ is the energy density and
$J^i$ the current vector. An initial data set is said to be
\emph{maximal} if $K=0$. If in addition $K_{ij}=0$, then it is called
time symmetric.

We proceed now to discuss some aspects concerning the way the Einstein
constraints (\ref{Hamiltonian}) and (\ref{Momentum}) are solved,
putting some emphasis on the relation of these ideas to numerical
implementations ---a major review on general aspects of the constraint
equations can be found in \cite{BarIse04}; see also
\cite{Coo00,Pfe05,Gou07b} for a numerical counterpart. For concreteness we
shall concentrate our discussion on the vacuum constraint equations.

\subsection{Topology of the initial hypersurface} \label{s:topology}

As mentioned already in section \ref{asympt_Euclidean}, from a Cauchy
initial value problem point of view, the natural requirement to obtain
a spacetime describing an isolated body is to have an initial
hypersurface $\Sigma$ that has at least one asymptotically Euclidean
end ---see equation (\ref{asympt_Euclidean}).  The theory of
hyperboloidal hypersurfaces is less developed, although some of the
essential ideas of the asymptotically Euclidean setting can be retaken
---see again \cite{BarIse04,AndChrFri92,AndChr93,AndChr96,AndChr94}.
Initial hypersurfaces with more than one asymptotic end are said to
have a \emph{non-trivial topology}. Note however, that the
introduction of multiple asymptotic regions is not the only way of
having initial data sets with non-trivial topology. The introduction
of a wormhole connecting two points in a hypersurface would also
result in a non-trivial topology. This was firstly exemplified in
\cite{Mis60}; with the use of the Isenberg-Mazzeo-Pollack (IMP) gluing
techniques ---see section \ref{s:gluing}--- it is now possible to
construct this type of initial data in a systematic way.

Initial hypersurfaces with non-trivial hypersurfaces have been
routinely used in the mathematical literature as a way of manufacturing initial
data sets for black hole collisions. The simplest example of the
latter are the usual time symmetric initial data for the Schwarzschild
spacetimes in isotropic coordinates, which contain two asymptotic
regions connected by a so-called \emph{Einstein-Rosen bridge, throat}
or \emph{wormhole}. This example can be readily be generalised to
contain $n+1$ asymptotic regions rendering
the so-called \emph{Brill-Lindquist initial data}
\cite{BriLin63}. Another approach, which renders time symmetric initial
data on a hypersurface $\Sigma$ containing $2$ asymptotic regions and
$n$ \emph{wormholes} and can be interpreted as the data for $n$ black
holes ---the \emph{Misner initial data}--- was given in \cite{Mis63}.

A first step towards the construction of non-time symmetric data for
black holes was given by Bowen \& York in \cite{BowYor80}, where an
analytic solution of the momentum constraint in the conformally flat
case was presented, the \emph{Bowen-York second fundamental form} ---a
discussion of the general solution of the so-called \emph{Euclidean
  momentum constraint} can be found in e.g.  \cite{DaiFri01,BeiOMu96}.
The so-called \emph{Brandt-Br\"ugmann} \emph{puncture initial data}
\cite{BraBru97} can be in some ways regarded as the non-time symmetric
generalisation of the Brill-Lindquist data.  It is important to
mention that all the initial data sets discussed in this last
paragraph are conformally flat ---see section
\ref{s:conformal_Ansatz}.  But this is merely a simplifying assumption
and in no way essential.

The fall-off conditions for $\gamma_{ij}$ and $K_{ij}$ to be asymptotically
Euclidean have already been discussed in section \ref{s:asymflat} ---see
equation (\ref{asympt_Euclidean}). As in the discussion of boundary
conditions for the whole spacetime, the discussion of the fall-off
conditions on $\Sigma$ can be geometrically reformulated in terms of
a conformally compactified 3-dimensional manifold $\hat{\Sigma}$
---see \cite{BeiOmu91,BeiOmu94,Fri86a,Fri98a}. The compactification
procedure is, in this circumstance, analogous to that induced by the
introduction of stereographic coordinates to compactify $\Real^2$ to
render the 2-sphere $\Sphere^2$. From this point of view, the
compactified initial hypersurface $\hat{\Sigma}$ contains, say, $m$
singled out points $\{i_1,\ldots,i_m\}$ corresponding to the $m$
asymptotic ends of the \emph{physical} initial hypersurface. For
example, the standard Schwarzschild initial data render a
$\hat{\Sigma}$ with the topology of the 3-sphere $\Sphere^3$ and two
singled out points. The Brill-Lindquist data for 2 black holes would
render a $\hat{\Sigma}$ with 3 singled out points, while in case of
the Misner data the resulting $\hat{\Sigma}$ has the topology of a
torus with 2 singled out points.

From a mathematical point of view, working with $\hat{\Sigma}$
instead of $\Sigma$ has several technical advantages. It is simpler to
prove existence of solutions for an elliptic equation on a compact manifold
than on a non-compact one. The price one has to pay is that the equations will
be singular at the points at infinity $i_{1},\ldots,i_{m}$. However, these
singularities are mild. It is also simpler to analyse the fields in terms of
local differentiability in a neighbourhood of the points at infinity than in
terms of fall-off expansions at infinity in $\Sigma$. This approach to
discussing the theory of the constraint equations has been introduced and used
in various applications in \cite{Fri86a,Fri98a,Dai01a,Dai01b,Dai01c,BeiHus94}.

The \emph{puncture initial data} used in many of the simulations of
the coalescence of black holes, as introduced in
\cite{BraBru97}, is a sort of compromise between using the physical
hypersurface $\Sigma$ and the compactified one $\hat{\Sigma}$. For
punctured data, one of the asymptotic ends ---the
so-called reference, or physical asymptotic end--- is not compactified. The resulting
initial hypersurface $\hat{\Sigma}$ will have the topology of
$\Sphere^3 \setminus i_1$. The singular behaviour of the constraints 
near the punctures reflects in a singular behaviour of their
solutions that is mild and
well understood ---it is related to the singular behaviour of the
Green functions of the elliptic operators appearing in the constraint
equations ---see \cite{BeiOmu91,BeiOmu94}.  Numerical relativists
interpret this phenomenon as a coordinate singularity at the puncture
---see \cite{BeiOmu91,BeiOmu94,HanHusBru06,HanHusPol07}. The
attractive of puncture initial data sets for numerical relativists is
that they represent black holes in $\Real^3$ without excision and it
is well understood how to construct puncture initial data for any
number of boosted, spinning black holes \cite{BraBru97,Dai03}.

The relevance for numerical relativists of initial data sets with
non-trivial topology stems from the fact that they allow to model
black hole spacetimes by only making use of the vacuum EFE: indeed, if
there is a wormhole in the initial data, then it can be concluded that
the spacetime will be geodesically incomplete \cite{Gan75}.  Note
however, that in all this argument one is indirectly assuming that
Cosmic Censorship holds and that the resulting spacetime is really a
black hole spacetime. In particular, from a physical point of view one
is not interested in what happens in the extra asymptotic ends ---see
however the discussions in section \ref{s:moving_punctures} on
\emph{why the puncture methods work}. From the point of view of a
physicist, it is fair to say that the use of initial data sets with
non-trivial topology is a trick ---there have been, however, some
works on trying to analyse the possible physical differences between
initial data sets with different topology but describing the same
system ---see e.g.  \cite{Giu90,Giu92,Val03b,SmaCadDeWEpp76,CanKul82}.

\subsection{The Conformal Method.} \label{s:conformal_Ansatz} 
A standard approach to study the space of solutions of the constraint
equations is the so-called conformal method ---or conformal Ansatz.  The method
not only provides a systematic approach to make statements about the
existence of solutions to the constraint equations, but also gives a
starting point for the numerical computation of the solutions to the
equations. A recent account of what is known about solutions to the
constraint equations is given in \cite{BarIse04} ---see also
\cite{Coo00,Pfe05,Gou07b}.

The conformal method is based on the conformal rescaling of the metric
\begin{equation}
\gamma_{ij} = \Psi^4 \tilde{\gamma}_{ij} \ \ ,
\end{equation}
and on the combined conformal rescaling and tensor splitting of the traceless
part $A^{ij}$ of the extrinsic curvature $K^{ij}$ 
into a \emph{transverse-traceless} and
\emph{longitudinal part}. Different ---non-equivalent--- constructions follow
from the non-commutativity of the conformal rescaling and tensor splitting
process. A key point in all of them is the need of a specific
rescaling for the (conformal) traceless part of the extrinsic curvature in
order to set the constraint equations in a form which is optimal for their
analytic study.
In particular, if $K$ is constant, the conformal method
reduces the problem of solving the constraint equations 
to a decoupled underdetermined elliptic problem.

Much has been learned about solutions of the Einstein constraint
equations using the conformal method. However, many other interesting
theoretical questions cannot be addresses using these methods. The
main problem being that the conformal method produces solutions of the
constraint equations, essentially from scratch. It turns out
that it is very difficult to encode physics into the \emph{conformal
  metric} $\tilde{\gamma}_{ij}$ and in the symmetric
transverse-traceless tensor $A^{TT}_{ij}$ ---which act as seeds of the
method. For example, there seems to be no way of constructing
systematically initial data for stationary solutions, although there
is an \emph{a posteriori} check for the case of time symmetric data
\cite{Dai04c}. Most of the natural simplifying assumptions that are
directly suggested by the structure of the conformal method happen to
have no essential physical or geometrical content ---most notably the
assumption of conformal flatness.  This is bluntly illustrated by
results showing that there are no conformally flat hypersurfaces in
the Kerr spacetime \cite{GarPri00,Val04b,Val04a}. Some examples of
solutions to the Einstein constraints where it has been possible to
put some desirable physical properties with some level of success can
be found in for example \cite{Dai01a,Dai01b,DaiLouTak02}. In these
last references either a non-flat metric or a different choice of the
Bowen-York Ansatz for the extrinsic curvature ---see
\cite{BowYor80}--- have been used.  Another approach to prescribe
physics is the use of tensors $\tilde{\gamma}_{ij}$ and $A_{ij}$
motivated by post-Newtonian analysis as in \cite{Nis06}.

It is noted that given the relation of $K_{ij}$ with the
 \emph{canonical momentum} in the \emph{Hamiltonian formulation} of the theory
 \cite{ArnDesMis62}
\begin{equation}
\pi^{ij}=\sqrt{\gamma}(K \gamma^{ij} - K^{ij}) \ \ ,
\end{equation}
this view of the constraints can be referred to as a \emph{Hamiltonian
  point of view} \cite{Yor99,PfeYor03}.  No reference to the slicing
is present and in particular no reference to the lapse function or
shift vector emerges in the formulation of the constraints.
This formulation reflects the character of the initial data problem 
 as a one-hypersurface embedding problem. In this context, in particular
in the numerical community, the conformal Ansatz is also referred to as
the {\em extrinsic curvature} approach.

\subsection{The conformal thin sandwich method.}
\label{s:CTS}
The idea of a thin-sandwich approach to GR ---in the spirit of the
Jacobi principle in Mechanics--- seems to have been firstly introduced
by Baierlein, Sharp and Wheeler (BSW) \cite{BaiShaWhe62}.  They
conjectured the possibility of specifying the spacetime metric
$g_{\mu\nu}$ on two hypersurfaces and then of recovering the spacetime
in between.  In particular, when applied to two infinitesimally closed
spatial slices ---the BSW Thin Sandwich (TS) conjecture--- this
approach would imply a manner of providing initial data for a Cauchy
evolution of the Einstein system. The TS conjecture, through its
two-hypersurfaces Ansatz, proposes a different perspective to the
constraint problem as compared to the one-hypersurface embedding of
the extrinsic curvature approach. In this setting one aims at
prescribing the induced metric and its time derivative $(\gamma_{ij},
\dot{\gamma}_{ij})$, in what can be seen as a \emph{Lagrangian}
formulation of the problem.  However, although this conjecture can be
seen to work for particular examples, the analysis in \cite{BarFod93}
shows that the conjecture fails in generic cases.

More recently ---see also previous works \cite{Ise78,IseNes80,WilMat89,CooShaTeu96} and section
\ref{s:helical}--- a new interpretation of the Einstein 
constraint equations, partly on the spirit of the TS view point, has
been given in \cite{Yor99}. In this approach, the freely specifiable
data on the initial hypersurface $\Sigma$ consists of a conformal
metric $\tilde{\gamma}_{ij}$ and its \emph{time variation}
$\tilde{u}_{ij}\equiv \dot{\tilde{\gamma}}_{ij}$ ---which is imposed
to be traceless. The latter can be shown to amount to the choice of
the conformal metric in two infinitesimally close slices. Using the
kinematic relation between the time derivative of the physical
3-metric and the extrinsic curvature, the traceless part $A_{ij}$ can
be written explicitly in terms of the lapse function $N$ and shift
vector $\beta^i$ ---cf. notation in section \ref{s:3+1}.
A function $\tilde{N}$ conformally related to the
physical lapse is identified as the naturally freely specifiable
parameter.  This conformal rescaling together with the absence of
rescaling of the shift vector, permits to {\em derive} the key
conformal rescaling for $A_{ij}$ introduced by Lichnerowicz \cite{Lic44}
and leading to the
decoupling of Hamiltonian and momentum constraints in the constant $K$
case.  The specification of the free data
$(\tilde{\gamma}_{ij}, \dot{\tilde{\gamma}}_{ij}; K, \tilde{N})$
transforms the constraints into an elliptic system on $\Psi$ and
$\beta^i$ similar in form to the one obtained in the {\em extrinsic
  curvature} approach \cite{Yor99}. In this \emph{Conformal TS
  approach} (CTS) the role of every part of the
metric is explicit. The formulation has been extended and improved in
\cite{PfeYor03}, which in particular reconciles the {\em
  Hamiltonian} and {\em Lagrangian} points of view. The new key
feature is the introduction of a weight function $\sigma$ which
permits to recover the commutativity between conformal rescalings and
tensor splittings. Moreover, with the choice $\sigma=2N$, the vanishing
of the tranverse-traceless part of the extrinsic curvature
$\tilde{A}_{TT}^{ij}$ characterises stationarity, in agreement with
the general (intuitive) understanding of $\tilde{A}_{TT}^{ij}$ as the
piece encoding the radiative ---or dynamical--- degrees of freedom.
Regarding the free data, the approach in \cite{PfeYor03} pushes
forward the \emph{Lagrangian analogy}. Understanding $K$ as a
configuration variable, it is proposed the prescription of $\dot{K}$
instead of $\tilde{N}$ at the (first) initial slice.  This translates
into a fifth elliptic equation for the conformal lapse $\tilde{N}$
which together with the CTS equations gives rise to the {\em Extended
  Conformal Thin Sandwich} (XCTS) system ---see e.g.
\cite{PfeYor03,Pfe05,BauOMuPfe07} for the list of the equations.

One of the advantages of the CTS approach is that using it, it seems
easier to prescribe physics than in the standard approach based on the
{\em extrinsic curvature} approach. In particular, as discussed in
\cite{Coo00,Gou07b}, the CTS method motivates definite Ans\"atze for
the free part in the construction of stationary initial data ---at
least in the axisymmetric case.  More generally, this CTS approach has
been used also to construct binary initial data sets in
\emph{quasi-equilibrium} ---see e.g.
\cite{BauCooSch97,BonGouMar99,GouGraTan01,UryEri00} or Ury\=u's IHP talk
\cite{IHP} for the case of neutron stars;
\cite{GouGraBon02,GraGouBon02,YoCooSha04,CooPfe04,CauCooGri06} for the
case of binary black holes; and \cite{BauSkoSha04,TanBauFab05,Gra06}
for the mixed black hole-neutron star binary systems.

\subsubsection{Uniqueness issues in the XCTS system.} 
\label{s:uniqueness_XCTS}
In contrast to the CTS and closely related
elliptic systems ---see e.g. \cite{OMuYor73,OMuYor74,York79} 
and also further references in \cite{Yor99,PfeYor03}---
there is to date a lack
of systematic analyses of the mathematical properties of XCTS
system. In particular, no definite results are available about
existence conditions in this system, and very little is known about
uniqueness. Numerical evidence of the non-uniqueness of solutions of the XCTS elliptic system has been found 
\cite{PfeYor05}. This evidence is of relevance as the XCTS system or
somewhat related ideas are used to construct initial data for binary
neutron star and black hole binary systems ---cf. the literature
listed in the previous paragraph. Moreover, it also has implications
for the groups implementing constrained evolution schemes  ---cf.
section \ref{s:mixed_ell-hyp}.

A first analysis of the non-uniqueness issue,
looking in particular at the question of genericity, has been given in
\cite{BauOMuPfe07}. The most complete analysis available of the
non-uniqueness of the solutions to the XCTS system, using
\emph{Lyapunov-Schmidt bifurcation theory}, has been given in
\cite{Wal07} ---see also Walsh's IHP talk \cite{IHP}. It is found that if the
linearised system has a Kernel for sufficiently large data, then there
exists a \emph{parabolic branching} of the solutions as the one
happening in \cite{PfeYor05}. The prosaic reason for the existence of the
non-trivial Kernel is the presence of a \emph{wrong} sign in one of
the equations of the XCTS system ---more precisely, in the equation
for the lapse. Thus, standard methods based on the use of a
\emph{maximum principle} cannot be employed. The numerical evidence
shows that the assumption of a nontrivial Kernel is certainly not
far-fetched, and actually occurs in numerical implementations. The situation
gets even worse if a treatment of boundary conditions is required
---see e.g. \cite{JarAnsLim07}.

The issue of the non-uniqueness of solutions to the XCTS equations
is an example
of the tension underlying the relation between numeric and
mathematical relativists: a successful method of prescribing physics into
numerical simulations with non-desired
mathematical properties. In the absence
of better alternatives, the numerical community seems prepared to take
the risk.

\subsection{Construction of initial data sets using the Isenberg-Mazzeo-Pollack gluing.} \label{s:gluing}

In recent years it has been possible to address a number of issues concerning
properties of initial data sets using a tool of geometric analysis: gluing 
---cf. Pollack's IHP talk.
While the conformal method is a procedure to construct solutions to the
constraint equations from scratch, gluing constructions allow to obtain new
solutions by means of direct sums of two already known solutions. Analytic
gluing techniques have played a prominent role in many areas of differential
geometry: some notable applications include the study of the topology of
4-manifolds and the study of minimal and constant mean curvature surfaces
in Euclidean 3-space.

In \cite{IseMazPol02,IseMazPol03,ChrIsePol04} a particular gluing construction
for solutions to the Einstein constraint equations was developed under the
assumption of a constant $K$ ---the
\emph{Isenberg-Mazzeo-Pollack (IMP) gluing}. This construction allows to
demonstrate how spacetimes can be joined by means of a geometric connected
sum, or how a wormhole can be added between two points in a given spacetime
---at the level of the initial data. This construction makes it possible to
address a number of issues concerning the relation of the spatial topology to
the geometry of solutions of the constraints and the constructibility of
multi-black hole solutions \cite{ChrMaz03}.

The sharpest possible gluing theorem for the Einstein constraint equations has
been developed in \cite{ChrIsePol04}. Using it, it is possible to show that for a generic solution of the constraint equations and
any pair of points in this solution, one can add a wormhole connecting these
points to the solution with no change in the data away from a neighbourhood of
each of these points. Further, one can show that for almost any pair of initial data sets ---including, say, a pair of black hole data sets, or a cosmological data set
paired with a set of black hole data--- one can construct a new set which joins
them.

In order to assess the relevance of the IMP gluing construction for
numerical relativity, we present a rough overview on how the IMP
gluing method works.  Consider two initial data sets
$(\Sigma_1,\gamma_{ij}^1,K^1_{ij})$ and
$(\Sigma_2,\gamma^2_{ij},K^2_{ij})$ for the EFE.  Let $p_1\in
\Sigma_1$ and $p_2\in \Sigma_2$ be two arbitrary points. The idea of
the IMP construction is to find an initial data set
$(\Sigma_{12},\gamma^{12}_{ij},K^{12}_{ij})$ such that $\Sigma_{12}$
has the topology of $\Sigma_1\# \Sigma_2$, where $\#$ denotes
\emph{the connected sum of manifolds} \footnote{The connected sum is
  performed by removing $\mathcal{B}(p_1)=\{\mbox{ball around } p_1\}$
  and $\mathcal{B}(p_2)=\{\mbox{ball around } p_2\}$ and then joining
  $\mathcal{S}_1\setminus \mathcal{B}(p_1)$ and
  $\mathcal{S}_2\setminus \mathcal{B}(p_2)$ at $\partial
  \mathcal{B}(p_1)$ and $\partial \mathcal{B}(p_2)$ with a cylinder,
  $\mathcal{I}$, of topology $\partial\mathcal{B}(p_1)\times [0,1]$.}.
The pair $(\gamma^{12}_{ij},K^{12}_{ij})$ is such that on
$\Sigma_1\setminus \{ \mbox{ball around } p_1\}$ it coincides with
$(\gamma^1_{ij},K^1_{ij})$ and on $\Sigma_2\setminus \{ \mbox{ball
  around } p_2\}$ with $(\gamma^2_{ij},K^2_{ij})$.

In order to construct the pair $(\gamma^{12}_{ij},K^{12}_{ij})$ satisfying the
Einstein constraints, one considers a \emph{conformal} metric
$\hat{\gamma}^{12}_{ij}$ with the property that on $\Sigma_1\setminus
\mathcal{B}(p_1)$ and $\Sigma_2\setminus \mathcal{B}(p_2)$ coincides with
$\gamma^1_{ij}$ and $\gamma^2_{ij}$ , respectively, and on the cylinder
$\mathcal{I}$ it interpolates between them.  Similarly, one constructs a
$\hat{\gamma}^{12}_{ij}$-traceless tensor $A^{12}_{ij}$ which again
interpolates between $A^1_{ij}$ and $A^2_{ij}$. By construction, the resulting
pair renders a solution to the constraints on $\Sigma_1\setminus
\mathcal{B}(p_1)$ and $\Sigma_2\setminus \mathcal{B}(p_2)$ but not on the
cylinder $\mathcal{I}$. The machinery of the conformal method ---see section \ref{s:conformal_Ansatz}--- is used then to
find a solution to the constraints on $\Sigma_1 \# \Sigma_2$ using the pair
$(\hat{\gamma}^{12}_{ij},A^{12}_{ij})$ as seeds for the procedure. Some amount
of technical work has to be invested, once the solution
$(\gamma^{12}_{ij},K^{12}_{ij})$ has been obtained to guarantee that the
solution coincides with the original $(\gamma^1_{ij},K^1_{ij})$ and
$(\gamma^2_{ij},K^2_{ij})$ outside the gluing neighbourhoods
$\mathcal{B}(p_1)$ and $\mathcal{B}(p_2)$.

The conditions on the initial data sets to be glued by the afore-discussed
method seem,  at first sight, relatively mild. Essentially, it is only
required that the initial data sets are such that the domain of influence of
the neighbourhoods $\mathcal{B}(p_1)\subset \Sigma_1$ and
$\mathcal{B}(p_2)\subset \Sigma_2$ have no Killing vectors. The latter
condition can be reformulated in an intrinsic way in terms of the so-called
\emph{Killing Initial Data (KID)} ---see e.g. \cite{BeiChr97b}. More
precisely, the condition for gluings is that there are no KIDs in
$\mathcal{B}(p_1)$ and $\mathcal{B}(p_2)$ ---in particular, the latter implies
that there are no Killing vectors in the neighbourhoods. The condition on the
non-existence of KIDs has been shown to be generic in a very specific sense
---see \cite{BeiChrSch05}.

Every step in the procedure as described in, say, references
\cite{IseMazPol03,ChrIsePol04} is completely constructive. However, it
would require to solve elliptic equations on manifolds with nontrivial
topology.  This complication is aggravated by the fact that, because of
the requirement of the non-existence of Killing vectors in
the initial data sets to be glued, it is not possible to use initial
data sets with some symmetry ---like axial one. Furthermore, the
standard simplifying assumption of conformal flatness may complicate
matters further. The reason is that the only conformally flat initial
data with trivial topology is Minkowski data. Thus,
the use of conformally flat initial data to glue would require gluing
data which already have non-trivial topology.

\subsection{The Corvino-Schoen gluing method. }
\label{s:corvino_gluing} An alternative gluing method for the
construction of solutions to the constraint equations, which may be of
relevance for numerical applications, is the so-called
\emph{Corvino-Schoen (CS) gluing construction}. This method has been
developed in \cite{Cor00,CorSch06,ChrDel02,ChrDel03} and it allows to
smoothly glue any interior region of an asymptotically Euclidean
initial data to the asymptotic region of a slice of a stationary
spacetime ---e.g. Kerr initial data. For time symmetric asymptotically
Euclidean solutions of the constraints, this method glues any interior
region to an exterior region of a slice of the Schwarzschild
spacetime.

The procedure is best explained by considering the time symmetric
situation in which one tries to glue a time symmetric solution of the
constraints to an Schwarzschild exterior. As it is well known, for
time symmetric initial data sets the constraint equations reduce to
\begin{equation}
{}^{(3)}\!R[\gamma_{ij}]=0,
\end{equation}
where $[\gamma_{ij}]$ in the last expression highlights the fact that in what
follows the Ricci scalar will be regarded as a mapping between the spaces of
3-metrics on $\Sigma$ and scalars on $\Sigma$. Under certain circumstances
this mapping happens to be an isomorphism. That is, given a metric $\gamma_{ij}$ and a scalar $f$ on $\Sigma$ such that
\begin{equation}
{}^{(3)}\!R[\gamma_{ij}]=f,
\end{equation}
then if a further scalar field $g$ happens to be close enough to $f$ ---in a
functional sense--- then there is a metric $\gamma'_{ij}$ close to $\gamma_{ij}$
such that
\begin{equation}
{}^{(3)}\!R[\gamma'_{ij}]=g.
\end{equation}
The CS method makes use of precisely this property of the Ricci
scalar mapping. 

Let $\gamma_{ij}$ be a metric defined on a asymptotically
Euclidean manifold 3-manifold $\mathcal{S}$ satisfying the time
symmetric constraints. Consider a subset $\mathcal{K}\subset
\Sigma$ which is obtained by removing from $\Sigma$ one of
the asymptotic ends. The boundary of $\partial \mathcal{K}$ regarded
as a subset of $\mathcal{S}$ will be required to lie in the asymptotic
region and to have the topology of the 2-sphere. Next, introduce some radial
coordinate $r$ so that the locus of $\partial \mathcal{K}$ is given by
$r=r_0$, with $r_0$ a constant. It is noted that $\mathcal{K}$ is not
necessarily compact, as it could well be the case that there is another
asymptotic end inside $\mathcal{K}$. Denote by $\mathcal{A}$ the asymptotic end of time symmetric Schwarzschildean data. This asymptotic end is
characterised by four parameters. Namely the mass, and the location of
the centre of mass, $c^i$:
\begin{equation}
\gamma^S_{ij}=\left(1+\frac{m}{2|x-c|}\right)^4 \delta_{ij}.
\end{equation} 
As part of the gluing construction, one connects the regions
$\mathcal{K}$ and $\mathcal{A}$ by means of an annular region
$\mathcal{C}$ to obtain a new asymptotically Euclidean 3-manifold
$\Sigma'$. A positive definite symmetric tensor $\hat{\gamma}_{ij}$ is defined on the
resulting 3-manifold by requiring it to be identical to $\gamma_{ij}$ on
$\mathcal{K}$ and to $\gamma^S_{ij}$ on $\mathcal{A}$. On $\mathcal{C}$, it
is chosen so that it interpolates smoothly between $\gamma_{ij}$ and
$\gamma^S_{ij}$. Now, if ${}^{(3)}\!R[\hat{\gamma}_{ij}]$ is close enough 
to the constant
vanishing scalar field ---this
can be achieved by moving $\partial \mathcal{K}$, regarded as a subset
of $\Sigma$, suitably further into the asymptotic region--- the
result about the isomorphism of the Ricci Scalar mapping guarantees
that there is a small ---again in a functional sense--- tensor $\delta
\gamma_{ij}$ such that $\hat{\gamma}_{ij}+\delta \gamma_{ij}$ is a 3-metric and more
importantly
\begin{equation}
{}^{(3)}\!R[\hat{\gamma}_{ij}+\delta \gamma_{ij}]=0, 
\end{equation} 
on $\Sigma'$. The theorems ensuring the possibility of performing the
above construction allow to show that the support of $\delta \gamma_{ij}$ is
$\mathcal{C}$, so that $\hat{\gamma}_{ij}+\delta \gamma_{ij}$ coincides with
$\gamma_{ij}$ on $\mathcal{K}$ and with $\gamma^S_{ij}$ on $\mathcal{A}$.

The precise details of the gluing construction require the
linearisation of the Ricci scalar mapping. In order to obtain a
differential equation to be solved, one has to consider the
composition of this linearisation and its adjoint to render a fourth
order elliptic problem. The non-linear problem is then solved by a
Newton iteration method. It turns out that in a similar way to what happens
in the IMP gluing, the presence of KIDs in the gluing region
$\mathcal{C}$ is an obstruction to the construction. The situation
here is actually more delicate as the mere presence of
\emph{asymptotic KIDs} could be enough to cause problems. Note that as
one moves the gluing region further and further into the asymptotic
region, the problem could worsen. In order to get around these
complications, one uses the freedom in the choice of the Schwarzschild data one
is gluing to. Indeed, it has been shown in \cite{Cor00} that
by a clever choice of the parameters $m$ and $c^i$ one can get around
the problems with the asymptotic KIDs. Unfortunately, it seems that precisely
this procedure, is non-constructive. Thus, a numerical implementation of the
CS gluing method would have to devise an alternative way of determining
the parameters of the Schwarzschild exterior. 

Although the above discussion has been limited to time symmetric
initial data sets which are to be glued to a Schwarzschild exterior,
it has been shown that generic metrics can be glued to a wide class of
exteriors ---including stationary ones. In this case, instead of just
considering the Ricci scalar mapping, one has to consider the general
\emph{constraint mapping} which sends a pair of symmetric tensors on
$\mathcal{S}$ to a scalar and a 3-vector ---see \cite{ChrDel03,CorSch06}.

The gluing constructions for initial data sets are remarkable
geometric and analytical results which have provided new insights into
the structure of solutions to the Einstein constraint equations.
Important to analyse and discuss is the hypothetical use of these
constructions in the simulation of black hole spacetimes. There are
various issues at stake here: of course one has the conceptual ones;
but there are also the computational aspects. Is it possible to
calculate everything one requires? As mentioned above, a
straight-forward implementation of the CS gluing technique may not be
possible as it contains non-constructive aspects. Nevertheless, the
existence of the gluing constructions has, at least, an indirect
connection with the discussion of the robustness of the simulations of
black hole spacetimes. In theory, one could use the CS techniques on
the Brill-Lindquist (BL) initial data, and construct from it a new
initial data set which will be exactly BL inside a compact set, but in
addition it will be exactly Schwarzschildean in the reference
asymptotic end. As already mentioned in section \ref{s:asymflat}, a
set of conserved constants for the Schwarzschild spacetime
---the Newman-Penrose constants--- are zero. On the other hand, for BL data
these are non-zero ---see \cite{DaiVal02}. As these constants are
conserved along null infinity, what one finds is that each spacetime
will have a different radiation content close to timelike infinity.
This could be interpreted saying that each spacetime approaches in a
different way the asymptotic state ---i.e. the Schwarzschild
spacetime.

\section{Evolution formalisms} 
\label{s:evol_formalism}
As discussed in section \ref{s:gauge_red}, different strategies in the
{\em gauge reduction} process lead to distinct evolution formalisms. A
natural classification of evolution formalisms in GR is according to
the type of initial value problem they solve. Following the discussion
in section \ref{s:diff_IVP}, they can be broadly classified as being
\emph{Cauchy, hyperboloidal} or \emph{characteristic}. In this review
we focus our attention on the Cauchy-type formalisms and will briefly
touch on some aspects of hyperboloidal and characteristic schemes.

Following reference \cite{BonLedPal03} we will distinguish between
evolution formalisms, understood as the schemes devised to deal with
the reduction process ---involving e.g. the choice of variables or the
strategy of resolution of the constraints--- and {\em evolution
  systems}, as the concrete PDE systems set in a particular gauge and
actually explicitly solved during the evolution.  PDE evolution
systems must include the reduced (or main evolution) system of section
\ref{s:gauge_red}, but can also incorporate additional PDEs ---e.g.
fixing the gauge system.  In general terms, a line can be drawn
between i) formalisms addressing specific mathematical issues ---such as
existence, uniqueness, completeness--- and ii) schemes aiming
at the numerical construction of explicit solutions.  We focus on some
evolution formalisms where interaction between analysts and numerical
relativists has been, or is expected to be, particularly fruitful.
Excellent reviews on different aspects of this problem can be found in
the literature
\cite{FriRen00,ShiYon02,LehReu04,Fri05,Reu98,Reu04,ShiYon02,Pre07}.
The following two sections are devoted to Cauchy schemes: manifestly
covariant schemes in section \ref{s:GH} and to the \emph{so-called}
$3+1$ formalisms in section \ref{s:3+1}. Non-Cauchy approaches are
considered in section \ref{s:other}.

\subsection{Cauchy manifestly covariant schemes: generalised harmonic 
(or general wave gauge) formulations and $Z4$ formalism}
\label{s:GH}

A classical approach to the study of the EFE consists in 
casting it 
as a wave equation for which there exists a well developed theory. 
Using the deDonder expression for the Ricci tensor \cite{DeD21,Foc59} and 
keeping in the left hand side only the principal part, equation (\ref{e:EFE})
is written as  
\bea
\label{e:Harmonic_evol}
-\frac{1}{2}g^{\rho\sigma}\partial_\rho \partial_\sigma g_{\mu\nu}  + 
\partial_{(\mu} \Gamma_{\nu)} = S_{\mu\nu} \ \ ,
\eea
where 
$\Gamma^\mu\equiv g^{\rho\sigma}\Gamma^\mu_{\rho\nu}$ and
$S_{\mu\nu}$ does not contain second derivatives of the metric.
Prescribing the gauge  $\Gamma^\mu=0$, equation (\ref{e:Harmonic_evol})
becomes a wave equation. Since $\Box x^\mu=-\Gamma^\mu$ (with $\Box$ 
denoting the scalar wave operator associated with the metric $g_{\mu\nu}$)
this choice corresponds to the use of {\em harmonic coordinates}.
These coordinates were early used in the first results on well-posedness
\cite{Cho52,FisMar72}. More recently, the elucidation
of the role of $\Gamma^\mu$ \cite{Fri85} 
---see also \cite{Gar02}--- led to the introduction of
general {\em gauge source functions} $H^\mu$, leading to 
{\em Generalised Harmonic} (GH) or {\em general wave gauge} conditions
\bea
\label{e:gauge_system}
\Box x^\mu=-\Gamma^\mu = -H^\mu \ \ .
\eea
In the language of section \ref{s:gauge_red}, this constitutes 
the {\em gauge system} in this covariant evolution formalism, 
and the {\em reduced system} follows from inserting (\ref{e:gauge_system}) 
into equation (\ref{e:Harmonic_evol}). The local Cauchy problem is well-posed
for this reduced system when appropriate initial conditions
consistent with the constraints (\ref{Hamiltonian})-(\ref{Momentum}) 
are employed 
---see e.g. \cite{FriRen00} and references therein for details and 
rigour. Consistency between the reduced and gauge systems ---namely
propagation of the gauge condition $Q_\mu\equiv \Gamma_\mu - H_\mu=0$
along the evolution--- is controlled by the {\em subsidiary system}. 
This follows from Bianchi identities, which imply the 
hyperbolic equation for $Q_\mu$
\bea
\label{e:subsidiary_system}
\nabla_\mu\nabla^\rho Q_\rho + {R^\rho}_\nu Q_\rho = 0 \ \ .
\eea
Imposing the gauge condition $Q_\mu=0$ and the
constraints (\ref{Hamiltonian})-(\ref{Momentum}) on the initial slice $\Sigma$
---which imply $\partial_t Q_\mu=0$, cf. \cite{FriRen00}--- 
the vanishing of $Q_\mu$ along the evolution and the overall
consistency are guaranteed.

This {\em subsidiary system  issue} exemplifies 
the different nature of the problems an analyst and 
a numerical relativist must address. 
In the analytical setting, the reasoning above shows
the consistency of the whole scheme. In particular
one no longer needs to consider the 
gauge system (\ref{e:gauge_system}) while studying
the properties of the reduced system. 
The situation gets more complicated in the numerical
setting, since numerical errors trigger constraint violation modes 
that invalidate the analytic argument ---see also \cite{Fri05,Fri05b}
for a discussion of the consequences of the intrinsic non-linearities
in equation (\ref{e:subsidiary_system}). The $Z4$ formalism 
proposed by Bona \emph{et al.} \cite{BonLedPal03,BonLedPal04,BonLedPal05,BonLehPal05} 
is particularly interesting in this context. Developed as 
a full spacetime generalisation of previous numerically motivated 
3+1 schemes \cite{BonMas92,BonMasSei95,BonLedPal02}, its  manifest covariant formulation
permits a general analysis in the lines of the harmonic schemes.
The key elements in this formalism are the introduction of 
a spacetime form $Z_\mu$ as a new dynamical field, and the
modification of the EFE to 
\bea
\label{e:Z4}
R_{\mu\nu}+ \nabla_\mu Z_\nu +\nabla_\nu Z_\mu = 8\pi\left(T_{\mu\nu}
  - \frac{1}{2}Tg_{\mu\nu}\right) \ \ .  
\eea 
The solution space of
this system extends that of GR, which is recovered for $Z_\mu=0$ ---in
fact, the EFE is recovered under the milder (non-generic) condition of
$Z_\mu$ being a Killing vector. The evolution of the new field is
driven by the Hamiltonian and momentum GR constraints
(\ref{Hamiltonian})-(\ref{Momentum}). Consequently, the $Z_\mu$ field
can be used as a measurement of GR constraint violations ---see
section II.B. in \cite{BonLedPal03} and \cite{BonLedPal05} for a
discussion involving the Bianchi identity and the constraint structure
of the theory.  As a further and key development for numerical
applications, reference \cite{GunMarCal05} introduces additional
damping terms in the Z4 system ---and also in the GH scheme--- to drive
the solution towards the GR solution submanifold, $Z_\mu=0$.

Together with the early and ongoing application in analytic studies of
GR equations ---see e.g. the work of Lindblad \& Rodnianski in section
\ref{s:asymflat}, and also the Kerr uniqueness proof by Klainerman
\& Ionescu discussed in section \ref{s:rigidity}--- these covariant
formalisms have led to evolution systems with a successful application
in numerical implementations.  Harmonic coordinates have been employed
in numerical studies of the IBVP in GR
\cite{SziSchWin02,SziWin03,BabSziWin06} and related evolution issues
\cite{BabSziWin06,BabSziWin06b}, or in the study generic singularities
\cite{Gar02} ---the latter also discusses GH coordinates.  In the
current numerical relativity scenario, it is specially relevant the
evolution of binary black holes through the merger with gravitational
wave extraction by Pretorius \cite{Pre05}, for which he employed a
generalised harmonic scheme \cite{Pre06} ---see also \cite{Pre07}---
using the damping terms based on the scheme in reference
\cite{GunMarCal05}. First-order symmetric hyperbolic forms of the
generalised harmonic codes have been developed in \cite{LinSchKid06}
---cf. also \cite{FisMar72,Alv02}--- where it is studied the
appropriate choice of gauge source functions $H^\mu$ for preserving
hyperbolicity of the whole evolution system. The latter includes now
not only the reduced system, but also the $H^\mu$ evolution equations.

\subsection{Cauchy 3+1 formalisms} \label{s:3+1} 

The Cauchy IVP involves the use of 3+1
spacetime spatial slicings. This applies in particular to the case of
the harmonic systems in the previous section.  We now consider a
family of evolution formalisms developed in the {\em $3+1$ framework} of GR
\cite{Darmo27,Lic44,Cho52,ArnDesMis62,Wheel64}, which provides the 
natural setting for the canonical or Hamiltonian
approach to GR ---notably inspired by GR quantisation
considerations \cite{Bergm56,Dirac58,ArnDesMis62,DeWit67,Ash91,Thi07}---
and is very extended in its 
application to numerical relativity
---e.g. \cite{York73,SmaYor78,York79,Gou07a}.
In this $3+1$
framework, in addition to the splitting of spacetime in terms of a
foliation by spatial hypersurfaces $(\Sigma_t)$, the basic variables
are fields naturally {\em living} on those slices $\Sigma_t$. In this
section, we use the expression ``$3+1$'' in this restricted sense,
customary in numerical relativity.  We denote the lapse function,
measuring the proper distance between slices $\Sigma_t$ in the normal
$n^\mu$ direction, by $N$ and the shift vector, governing the choice
of coordinate system on each slice, by $\beta^\mu$.  The {\em
  evolution vector} is then written as $t^\mu = N n^\mu + \beta^\mu$.
In this $3+1$ decomposition, the EFE splits in two sets. The first one
consists of the constraints in equations
(\ref{Hamiltonian})-(\ref{Momentum}) whereas the second one ---together
with the kinematic relationship between the 3-metric $\gamma_{ij}$
and the extrinsic curvature $K_{ij}$--- is frequently referred to as the
ADM evolution system (but is actually due to 
York \cite{York79}): \bea &&\left(\partial_t - {\cal
    L}_\beta\right)\gamma_{ij}= -2N K_{ij}
\ \ , \nonumber \\
&&\left(\partial_t - {\cal L}_\beta\right) K_{ij} =
-D_iD_jN  \nonumber \\
&&\hspace{0.7cm}+N\left({}^{(3)}\!R_{ij}+ K K_{ij} - 2{K_i}^k K_{kj}
  +4\pi\left[(S-E)\gamma_{ij} - 2S_{ij}\right]\right).
\label{e:ADMsystem} \eea The ADM system is not uniquely defined, since
terms homogeneous in the constraints
(\ref{Hamiltonian})-(\ref{Momentum}) can be added. Moreover, it
becomes an actual PDE evolution system once prescriptions for the
lapse $N$ and shift $\beta^i$ are actually supplied.  The resulting
evolution systems have suffered from the presence of numerical
instabilities and the closely related violation of the constraints.  A variety
of 3+1 formalisms have been developed along the years with the aim of
coping with these problems ---see \cite{ShiYon02} for a (non-updated)
review. Depending on the strategy adopted to handle the constraints,
evolution formalisms can be broadly classified as: i) {\em free
  evolution formalisms} if constraints
(\ref{Hamiltonian})-(\ref{Momentum}) are not enforced during the 
evolution, and ii) {\em constrained evolution formalisms}, that can be {\em
  partially-} or {\em fully-constrained}, according to the
prescription of some or all the constraints, respectively.

The underlying {\em intrinsic hyperbolicity} \cite{Fri05} of the EFE is an
ingredient exploited in the task of obtaining a specific PDE evolution
system. However, this does not mean that all the resulting evolution
systems are necessarily hyperbolic.  A clear example is provided by
constrained schemes that, due to the elliptic nature of the
constraints, lead to mixed elliptic-hyperbolic systems ---see however
\cite{GenGeoKhe86}. Among the many possible criteria, in the present
general discussion we will classify 3+1 evolution systems according to
the presence or absence of elliptic equations in the system.

\subsubsection{Evolution systems not containing elliptic equations.}
A significant effort has been devoted to casting the EFE ---under the
appropriate gauge conditions--- as a PDE hyperbolic system. The reason
behind is the analytic control these systems provide on existence,
uniqueness and stability issues \cite{Reu98,Reu04} ---see LeFloch's IHP talk for
some examples illustrating the richness of this approach in the 
interplay between non-linear PDEs and geometry.  This is the case
of strongly hyperbolic systems ---with their well-posedness results on
the IVP--- and of symmetric-hyperbolic systems, where energy estimates
enhance the analytic control particularly in the IBVP permitting the
use of maximally dissipative boundary conditions.  In the generalised
harmonic formalisms discussed in the previous section \ref{s:GH}
hyperbolicity is manifest.  With regards to systems derived from the
ADM formulation \cite{ShiYon02}, hyperbolicity analysis is more
complex due to the loss of covariance.  This also applies to certain
3+1 formalisms, that can be derived from covariant formalisms by
performing a 3+1 splitting of the fields ---e.g. 3+1 versions of the
$Z4$ scheme in \cite{BonLedPal03}.  A first class of systems in the
3+1 context consists of first order ---in time and space--- evolution
systems. Among these, the KST system \cite{KidSchTeu01} provides a
very general class from which, for example, the Fritelli-Reula
\cite{FriReu94,FriReu96} or the Einstein-Christoffel \cite{AndYor99}
systems can be recovered. The Bona-Mass\'o systems
\cite{BonMas92,BonMasSei95} include some additional variables
---expressible in terms of the connection coefficients of the three
metric, as the BSSN system below--- and can be related by variable
reduction to KST systems \cite{BonLedPal02}.  A second class of 3+1
systems has been studied in which the second-order-in-space
aspect of the ADM system is explicitly maintained. 
In this sense, reference \cite{GunMar06b} presents a
very general second-order evolution system that includes the
successful BSSN \cite{ShiNak95,BauSha98} or the NOR \cite{NagOrtRer04}
systems.  In this context of second-order systems, we also highlight
the work in \cite{BonLedPal02}, as a predecessor of the $Z4$ formalism
discussed in section \ref{s:GH}. We refer the reader to reference
\cite{ShiYon02} for the presentations of a third subclass, the {\em
  asymptotically constrained} systems aiming at the construction of an
evolution system where the constraint surface is an attractor ---cf.
in particular the discussion of the {\em $\lambda$-system} and {\em
  adjusted system} approaches.  Many evolution systems can be obtained from
the previous systems ---or closely related ones---  by adopting distinct
prescriptions for the lapse and shift.  An exhaustive account of all
the resulting 3+1 systems is beyond the scope of this article.
Assessment of analytic well-posedness ---strongly dependent of gauge
choices--- and of the relations between first and second-order
formulations, can be found in
\cite{BonLedPal02,NagOrtRer04,SarCalPul02,GunMar06b}.

\subsubsection{Mixed elliptic-hyperbolic systems.}
\label{s:mixed_ell-hyp}
Constrained formalisms are not the only avenue to mixed
elliptic-hyperbolic systems. There are geometric or physically well
motivated choices for $N$ and $\beta^i$ which are elliptic in nature
and thus lead to a mixed system. We present some of the
elliptic-hyperbolic PDE systems discussed in the literature:

{\em a) Free systems.}  They involve elliptic gauge choices for the
slicing and/or the shift.  An example of this is provided by the first
of the axisymmetric 2+1+1 systems considered in reference
\cite{Rin06}. A paradigmatic example is presented by Andersson and
Moncrief in \cite{AndMon03}, where the authors establish the
well-posedness of the IVP for a system containing elliptic equations
for the lapse and the shift. These equations follow, respectively,
from a constant mean curvature condition and a spatial harmonic gauge.
To the best of our knowledge, no numerical counterpart has been
implemented.  Other recent analytic studies can be found in references
\cite{ChoYor95,ChoYor96,ChoYor96b}.

{\em b) Partially constrained systems.}  The enforcement the
Hamiltonian constraint as an equation for a conformal factor
  has been widely used in axisymmetric codes since the
eighties \cite{BarPir83,StaPir85,Eva86}.  More recently, maximal
slicing and the Hamiltonian constraint have been employed in
\cite{GarDun01}, whereas the third constrained 2+1+1 system in
\cite{Rin06} implements maximal slicing and imposes the momentum
constraint ---all this, still in the axisymmetric setting.

{\em c) Fully constrained systems.}  Early systems of this kind were
implemented in 2D codes \cite{Eva89}, both for non-rotating
\cite{ShaTeu92} and rotating spacetimes \cite{AbrCooSha94}. More
recently, axisymetric 2+1+1 codes \cite{ChoHirLie03a,ChoHirLie03b}
have been used in the analysis of critical collapse. In this line,
reference \cite{Rin06} provides again an example of constrained system
of this type, where ill-posedness is concluded after a maximum
principle analysis of the involved scalar elliptic equations ---see
discussion in section \ref{s:uniqueness_XCTS}.  Regarding the full 3D
case, fully-constrained schemes have been discussed in
\cite{BonGouGra04,AndMat05} and Moncrief's IHP talk \cite{IHP}. The
fully-constrained scheme proposed by the Meudon group
\cite{BonGouGra04} makes use of a 3+1 conformal decomposition of
spacetime based on slices reaching spatial infinity $i^0$. The
elliptic part of the PDE system includes equations for the lapse, the
conformal factor and the shift, following from maximal slicing $K=0$,
the Hamiltonian constraint and a combination of the momentum
constraint and a {\em Dirac-like} gauge, respectively.  Only two
hyperbolic scalar modes are evolved, and the rest of the components
are reconstructed by using the Dirac gauge ---cf. Novak's IHP
talk \cite{IHP}--- and a unimodular condition on the conformal metric.
The scheme has been numerically tested for gravitational wave
spacetimes. Regarding the scheme proposed in Moncrief's IHP talk
\cite{IHP}, it is ultimately devised to bypass the limitations in the
extraction of gravitational radiation at finite distances by using instead
the natural boundary of the problem ---i.e. future null infinity.
Even though important features are shared with \cite{BonGouGra04}
---use of a conformal 3+1 decomposition, resolution of Hamiltonian and
momentum constraints along the evolution, elliptic gauges for the
shift--- it also represents a sound shift. Namely, it is not a Cauchy
formalism but rather of hyperboloidal type: hyperboloidal 3-slices
reaching up to $\scri^+$ are chosen by means of a {\em constant mean
  curvature} $K$ condition, implemented as an elliptic equation for
the lapse. Finally, it is worth noting an interesting proposal within
this scheme for the determination of the conformal class
representative $\tilde{\gamma}_{\mu\nu}$ by solving an elliptic Yamabe
equation.

\subsubsection{Current status of 3+1 systems.}
Most numerical groups make use of codes based on $3+1$ formalisms
derived from the ADM scheme.  Codes based on harmonic formulations
have produced excellent results \cite{Pre05,SziPolRez07}, but their
use is still limited to a smaller part of the community.  Among the
3+1 formalisms, the mainstream is represented by free systems not
involving the resolution of elliptic equations, and they have provided
the longest lived evolutions ---in particular BSSN.  In comparison,
mixed elliptic-hyperbolic systems have offered limited applicability
or are still at a preliminary stage ---cf. \cite{BonGouGra04} and
Moncrief {\em et al.} scheme \cite{IHP}.  On the one hand, constrained
systems are in principle expected to provide a better control on
instabilities related to violation of the constraints. On the other
hand, well-posedness in mixed systems is difficult to establish and,
in particular, characteristic fields in the hyperbolic part are
difficult to determine since part of the dynamics is encoded in fields
solved through the (non-causal) elliptic part.  In addition,
non-uniqueness issues in the XCTS system discussed in section
\ref{s:uniqueness_XCTS} could have strong implications on the
well-posedness of some fully-constrained evolutions schemes enforcing
a maximal condition for the slicing.  The numerical consequences of
this non-uniqueness in the elliptic sector are still unclear and
behaviour near the critical value of the parabolic branching discussed
in section \ref{s:uniqueness_XCTS} can be very dependent on the
details of the numerical implementation.  As a significant example,
authors in reference \cite{RinSte05} have renounced to the use of
mixed systems in \cite{Rin06}, opting for a 2+1+1 version of the $Z4$
formalism. The lesson here contained should certainly not be
underestimated. At the same time, it should not preclude the
development of such an alternative line of research that can only
result in overall benefits.

\subsection{Other approaches to evolution} \label{s:other} 
We proceed
to discuss some alternative approaches to the Cauchy formulation. We
focus our attention on some aspects of characteristic evolutions and
later on a particular version of the hyperboloidal problem.

\subsubsection{Evolution formalisms based on a Characteristic approach to GR.}
The characteristic initial value problem presented in section
\ref{s:diff_IVP} provides an avenue to the numerical construction of
spacetimes ---alternative to the Cauchy approach--- that is
particularly well suited for the study of gravitational radiation.
Since the pioneer work by Bondi and Penrose
\cite{Bon60,BonBurMet62,Pen63} on the characterisation of
gravitational radiation in terms of null hypersurfaces, developments
based on the use of Bondi-Sachs metric \cite{BonBurMet62,Sac62c} have
resulted in successful numerical evolution ---e.g. \cite{BisGomLeh97}---
and radiation extraction ---e.g. \cite{BabSziHaw05}--- in
a number of non-trivial cases, including single black hole spacetimes.
An excellent account can be found in the article review by Winicour
\cite{Win05}. The main drawback of this approach is the need to cope
with caustics formation during the evolution. A second obstacle lies
in the difficulty of encoding physics.  Alternatively, mixed
Cauchy-characteristic formulations  \cite{Bisho93}
(Cauchy-characteristic matching)
offer a compromise between stable evolutions and accurate treatment of
gravitational radiation.  Full application to astrophysical scenarios
seems currently limited as compared to the standard Cauchy approach.
On the other hand, regarding applications in more geometric settings ---in
particular, the study of global solutions--- characteristic or mixed
Cauchy-characteristic approaches provide a stimulating framework
for the fruitful collaboration between geometers and numerical
relativists. In this sense, we highlight the ongoing
activity in this line of research ---see e.g.
\cite{Bisho05,ReiBisLai07,GomBarFri07} and references therein.

\subsubsection{Conformal field  equations.}
\label{s:conformal_equations}
 \footnote{The content of his section is essentially due to A. Zengino\u{g}lu. We we are thankful for his enthusiastic input.}
The \emph{conformally regular} approach in numerical relativity is based on
analytic studies by Friedrich \cite{Fri81a,Fri83,Fri86b} and started
with numerical studies by H\"ubner and Frauendiener
\cite{Hub99a,Hub99b,Hub01a,Hub01b,Fra98a,Fra98b,FraHei02,Fra00,Fra03}.
In this approach one solves numerically a hyperboloidal initial value
problem for the conformally regular field equations ---for reviews see
\cite{Fra03,Fra04,Hus03}. The use of hyperboloidal foliations is
promising as hyperboloidal surfaces combine advantages of Cauchy and
characteristic surfaces.  Instead of approaching spatial infinity as
Cauchy surfaces do, they reach null infinity which makes them suitable
for radiation extraction.  Contrary to characteristic surfaces, these
spacelike surfaces are as flexible as Cauchy surfaces and can be used
in numerical calculations within the 3+1 approach. A difficulty with
the conformally regular field equations is that equations are
significantly larger than usual formulations of Einstein equations.
Due to the large number of constraint equations, numerical errors
require a stronger control on constraint propagation properties of the
system.  As there is not enough numerical experience with the
equations, one cannot use established methods easily to deal with the
encountered instabilities.  Another difficulty is that the equations
include, among others, evolution equations for the conformal factor.
The representation of null infinity depends, in general, on the
solution, so that the numerical boundary does not coincide with the
conformal boundary. As a consequence, a numerical boundary treatment
outside the physical spacetime is required and the calculation of the
unphysical part of the conformal extension wastes computational
resources. While these problems are not of a principal nature, they
have made progress in the conformally regular approach difficult. The
conformally regular approach as described above has been based on the
hyperboloidal initial value problem. As such, spatial infinity is not
part of the computational domain and one cannot calculate the maximal
development of Cauchy data. This has various drawbacks. In the
hyperboloidal approach it is not clear how the cut of the initial
hyperboloidal surface at null infinity is related to timelike or
spatial infinity. One would also like to be able to relate asymptotic
quantities such as mass or momentum defined at null infinity to
corresponding quantities at spatial infinity. An alternative
conformally regular system was proposed by Friedrich that allows one
to construct a \emph{regular finite initial value problem at spatial
infinity} \cite{Fri98a}. The study of this problem led to analytic
results concerning the applicability of the Penrose proposal and
smoothness properties of null infinity
\cite{Fri04,Val04a,Val04d,Val04e,Val05a}.  A major advantage of the
underlying system, called the reduced general conformal field
equations, is that the representation of the conformal factor is known
a priori in terms of initial data and that the system consists mainly
of ordinary differential equations except the Bianchi equation
admitting a symmetric hyperbolic reduction. The finite initial value
problem at spatial infinity has been studied numerically in
\cite{Zen07}. The Cauchy problem could be solved for the
entire Schwarzschild-Kruskal solution including timelike, spatial and
null infinity \cite{Zen06}. Also, certain radiative spacetimes could
be studied near spatial and null infinity.  This approach is currently
the only approach that allows the numerical study of both spatial and
null infinity in a single finite picture so that one has, in
principle, numerical access to the maximal development of Cauchy data
and can calculate global spacetimes including their entire
asymptotics. A further development of this approach might give
valuable input from numerics to geometry as most open problems in
mathematical relativity concern global questions.

A valuable input from geometry to numerics is the idea of null
infinity. Having null infinity in the computational domain would solve
both the outer boundary problem and the radiation extraction problem
in numerical relativity. Unfortunately, this idea could not yet be
implemented in astrophysically motivated numerical calculations based
on the Einstein equations. The main difficulty is due to the
appearance of formally singular terms arising from conformal
compactification of the metric. A reasonable approach is to choose a
gauge for the Einstein equations in which each formally singular term
attains a regular limit in a spacetime that admits a smooth conformal
compactification at null infinity. The construction of such a gauge in
the context of the characteristic approach has been known for a long
time \cite{TamWin66}.  This method has been quite successful within
the characteristic approach.  The underlying coordinates, however, do
not allow the simulation of highly dynamical strong fields due to
formation of caustics in the light rays generating the coordinate
hypersurfaces \cite{Win05}. The implementation of this idea within the
3+1 approach has turned out to be exceptionally difficult numerically
as well as analytically. In view of the recent breakthrough within
numerical relativity, it seems clear that yet another formulation of
the Einstein equations to deal with this problem cannot be regarded as
a practicable solution. What is needed is a novel numerical treatment
of null infinity that does not alter the successful simulation of the
sources in the interior.

An attempt to use a common reduction of the Einstein equations to
study null infinity has been made in \cite{Mon00, And02}. A
scri-fixing gauge has been suggested in the context of a hyperboloidal
initial value problem for the ADM reduction of the Einstein equations.
In this gauge the spatial coordinate location of null infinity is
independent of time. Such a gauge has first been constructed in
\cite{Fra98b} in the context of a frame-based conformally regular
field equations. A similar approach is followed in Moncrief's IHP talk
\cite{IHP}. A scri-fixing gauge is based on a mixed elliptic-hyperbolic 
system. A constant mean curvature condition fixes the hyperboloidal foliation
  by an elliptic gauge condition for the lapse. In such attempts, an
  important question is how to fix the conformal factor. It has been
  shown that the representation of the conformal factor in terms of
  coordinates can be given \emph{a priori }\cite{Zen07}. In this case, the
  well-posedness of the scri-fixing gauge has been proven. Further,
  each formally singular term arising from conformal compactification
  attains a regular limit at $\scri$.  Preliminary numerical
  experiments suggest that the method has some promise to be used in
  astrophysically motivated calculations.  In \cite{Zen07b} scri-fixing
  gauges have been studied in Minkowski and Schwarzschild spacetimes.
  A discussion of the tail behaviour of Schwarzschild is used to
  demonstrate the astrophysical relevance of the notion of null
  infinity for numerical applications.

\subsection{Some specific issues regarding evolution formalisms}

\subsubsection{Outer boundaries and the Initial Boundary Value Problem.}
\label{s:outer_boundaries}
The treatment of infinity poses diverse geometrical, numerical and
physical challenges. More specifically, among the different manners of
coping numerically with this issue we can mention compactifications of
spatial infinity \cite{Pre05}, use of hyperboloidal slices together
with compactification of null infinity
---\cite{Hub99a,Hub99b,Hub01a,Hub01b,Fra98a,Fra98b,FraHei02,Fra00,Fra03},
Moncrief's IHP talk \cite{IHP} and previous section 
\ref{s:conformal_equations}--- or implementation of a
characteristic scheme.  Alternatively, one can remove infinity from
the problem and consider the evolution of the EFE in a region bounded
by an outer timelike boundary. This leads to the discussion of the
well-posedness of the IBVP presented in section \ref{s:IVP}. A part of
the community no longer regards at this issue as a fundamental problem
in practise \cite{NFNR07}, since current simulations can push outer
boundaries sufficiently far away.  However, the relevance of the topic
is in particular reflected in the quantity of works 
in the subject ---also manifest
in the number IHP talks \cite{IHP} dealing direct or indirectly with
this topic; cf. IHP talks by Buchman, Rinne, Tiglio and Winicour.
The work by Friedrich and Nagy \cite{FriNag99} provided the first 
formalism in which well-posedness has been fully shown.
However, its formulation in terms of tetrads and the Weyl tensor makes
it difficult to implement using standard numerical
techniques/infrastructures. 
More recently Kreiss \& Winicour \cite{KreWin06} have presented a system based 
on the harmonic formalism for which the IBVP is well-posed 
in the generalised sense, and the result has been extended to 
well-posedness in the classical sense in \cite{KreReuSarWin07}.
Generally speaking, symmetric hyperbolic
systems play a critical role in the analysis of the well-posedness in
the IBVP problem, through the use of maximally dissipative boundary
conditions for getting rid of possible constraint violations ---cf.
\cite{GunMarCal05} for an alternative approach making use of
point-like damping terms.  In this context, pseudo-differential theory
of strongly well-posed systems in the generalised sense and techniques
from semi-group theory have been used to analyse IBVP well-posedness
---cf. \cite{Stewa98,CalSar03,CalGun05,SarTig05,Pre05b,BonLedPal05,
  Rin06b,LinSchKid06,RinLinSch07} for some other recent
references on the subject.  Absorbing boundary
conditions have been intensely studied in the last years
\cite{NovBon04,BucSar06,BucSar07,RuiRinSar07} ---cf. the review \cite{Sarba07}.

\subsubsection{Gauge conditions.}
\label{s:gauge}
The discussion about numerically and physically appropriate choices of
coordinates is a vast topic.  Here, we limit ourselves to briefly
account for some recent developments.  First, regarding sources
functions $H^\mu$ in GH formalisms, choices must be done in such a way
that the hyperbolicity of the GH formulation is not spoiled.
References \cite{Pre05,Pre05b,Pre06} make use of a wave evolution
equation for the determination of $H^\mu$, but more research is needed
on this specific topic \cite{Pre07}.  In this sense, reference
\cite{LinMatRin07} has provided new gauge drivers preserving
hyperbolicity in GH systems, and including a wider class of conditions
motivated by successful 3+1 implementations. The further discussion of
stability as well as the study of intrinsically GH-motivated slicings constitute an
open line of research.  In a different setting, the appropriate choice
of coordinates has been crucial in the successful binary black holes
implementations using the BSSN formalism
\cite{CamLouMar06,BakCenCho06}. In particular, modifications of the
\emph{1+log} slicings and \emph{Gamma-driver} condition for the shift,
have proved to be fundamental ingredients in the {\em moving puncture
  approach} to black hole evolution, where the punctures are advected
in the integration domain ---cf. section \ref{s:moving_punctures}. An
analysis of the well-posedness of the resulting evolution system has
been carried out in \cite{GunMar06b}, and in particular has given
indications of a breakdown for large shifts.  This binary black hole case
shows the relevance of using symmetry seeking coordinates
\cite{GarGun99}. Recent contributions in this sense are the {\em
  almost-Killing} condition derived from a variational principle in
\cite{BonCarPal05,BonLehPal05}, and its compatibility with the
adoption of singularity avoiding coordinates in schemes not
implementing excision \cite{BonAli07}.

\subsection{Final discussion on evolution formalisms}

Research in evolution formalisms for the Einstein equation provides a
fruitful area for the collaboration between mathematical and numerical
relativists.  However, the analytic issues on which we have focused in
this section should not shadow the full complexity of the problem, in
particular the fact that current numerical successes are a direct
consequence of decades of steady and systematic numerical
experimentation. The most successful codes ---e.g. generalised
harmonic or BSSN--- represent a compromise between well-posed but
complicate schemes ---like the Friedrich-Nagy system--- and formalisms
easy to implement but potentially ill-behaved, like the ADM approach.
The situation could be deceiving for the analyst, but it is just a
reflection of the overall complexity here involved. Analytic
well-posedness is not a guaranty of numerical stability ---cf.
discussion in \cite{BabSziWin06}--- but rather a necessary condition
for numerical convergence, once {\em consistency} of the scheme and
numerical stability are satisfied.  In sum, {\em analytic} issues are
only one aspect in the problem of choosing the appropriate numerical
evolution scheme. Together with them, the numerical relativist must
cope with {\em computational} and {\em physical} issues
\cite{BabSziWin06}.  In particular, a presentation of the formalisms
employed in numerics cannot be complete without a discussion of the
numerical techniques involved, a task that would require a full review
in itself.

For all these reasons, the discussion about evolution formalisms is
perceived by a good part of the numerical community as an essentially
{\em technical} discussion that, in a good measure, has lost its
critical character after the breakthroughs in the binary black hole
problem.  After decades of struggle, analytic and computational issues
are considered to be under reasonable control and, following this line
of thought, focus should consequently be shifted to the more physical
aspects of the problem ---cf.  special {\em New Frontiers in Numerical
  Relativity } issue \cite{NFNR07} and in particular the transcription
of the final discussion.  As far as {\em astrophysical} aspects in
numerical relativity are concerned, this attitude seems to be an
appropriate one ---although surprises should be expected when
addressing other physical problems.  However, regarding the
application of numerical techniques to deepen our understanding of the
geometric structures of GR ---what we have called {\em geometric
  numerical relativity} in section \ref{s:NRvsRA}--- it seems
reasonable to consider that further research in this subject is not
only fully justified, but will probably be needed as new geometric
and/or physical goals are set.

\section{State of the art of black holes} \label{s:art_bh}

If one wants to make use of numerical methods to calculate dynamical
black hole spacetimes then, in one way or the other, one has to resort
to an \emph{initial value formulation} ---see section \ref{s:IVP}.  It
is not obvious at all that an initial value formulation of the merge
of two black holes is the most adequate way of discussing this
problem\footnote{We thank S. Dain for discussion on this point.}
---one is dealing with an idealised astrophysical problem whose
initial conditions can not be set in a laboratory. The whole approach
hinges very heavily on the assumption that there are certain
\emph{robust aspects} of the output of the numerical simulations
---essentially the wave forms--- which are not dependent on the detailed way
the initial data are constructed. The only real dependence of the wave
forms should come from the physical parameters of the problem
---again, the masses and spin of the black holes--- but not from
things like the initial separation ---as the holes rather approach
from infinity undergoing an adiabatic process--- or if the initial
data is conformally flat or not.

The issue of the robustness of certain features in the simulation of
dynamical black hole spacetimes provides a rich arena for the
interaction between Geometry and Numerics. This is a very challenging
mathematical problem whose resolution probably will take long. For
the time being one will have to content oneself with the evidence
coming from the numerical simulations. Still, the assessment of this
evidence is in itself also a challenge, and may also provide further
ground for interaction.

In defence of the use of an initial value point of view, one can say
that it offers a framework to look at things. By framework it is being
understood a repertoire of mathematical techniques and theorems which
for example allows to state the well-posedness of the differential
equations governing the problem, so one knows that the problem has
been formulated in a consistent way. From a strict mathematical point
of view, it is not known if there exist spacetimes with a black hole
and gravitational radiation ---this is essentially the problem of the
nonlinear stability of Schwarzschild/Kerr, cf. the discussion in
section \ref{s:BHs}.

\subsection{Mathematical black holes}

\subsubsection{Rigidity and uniqueness of black holes.}
\label{s:rigidity}
A cornerstone of what it has been called \emph{the establishment's
  point of view on black holes} ---see e.g. \cite{Pen73}--- is the
evidence showing that the Kerr solution describes all stationary,
vacuum solutions to the EFE describing black holes. This uniqueness
assertion is also called a \emph{no hair theorem}. The relevance of
this result stems from the fact that it characterises all possible
asymptotic states of the general evolution of isolated systems in
vacuum. Thus, for the community of numerical relativists it justifies
the analysis of the late stages of black hole spacetimes by means of
perturbation theory on a Kerr background ---see e.g.  \cite{KokSch99}.

The problem of the uniqueness of black holes has been resolved
in two different manners. Starting from a stationary, asymptotically
flat black hole spacetime it follows that the stationary Killing vector
field must be tangent to the event horizon, $\mathcal{E}$, of the black
hole. If the Killing vector is null at the horizon ---i.e. it is a
\emph{Killing horizon}--- then it is possible to conclude that the
spacetime is actually static \cite{SudWal93}. For static black hole
spacetimes, the uniqueness results state that the spacetime has to be
a Schwarzschild spacetime ---see e.g. \cite{Chr93,Isr67,BunMas87}. If
the stationary Killing vector is spacelike on the horizon then from
Hawking's area law one can conclude the existence of another vector,
$\eta^\mu$, tangent to the generators of the horizon which is a
Killing vector on the horizon \cite{HawEll73,IseMon83,ChrDelGalHow01}.
This extra Killing vector field of the horizon can be used to define
the notion of \emph{surface gravity} which is of great relevance in
the discussion of the thermodynamics of black holes. In particular if
the surface gravity is non-zero, then the horizon is said to be
\emph{non-degenerate}. In the case of a non-degenerate horizon it is
customary to assume that the horizon is a smooth null hypersurface
consisting of two components ---a \emph{bifurcate horizon}---
$\mathcal{E}=\mathcal{E}^+\cup\mathcal{E}^-$ intersecting at a
2-surface with the topology of the 2-sphere ---this assumption is
supported by \cite{RacWal92}.

Given that it is known that the Kerr solution with $0\leq a=J/M \leq M$ is
the only stationary, axially symmetric, vacuum black hole with
non-degenerate and connected horizon ---see \cite{Car71,Rob75}---, it
is suggestive to try to extend the vector $\eta^\mu$ in the horizon to
a Killing vector of the region exterior to the black hole ---the
\emph{domain of outer communication}. The problem with this approach
is that it requires posing a boundary value problem on characteristic
hypersurfaces (the horizon of the black hole): one constructs a wave
equation for the vector $\eta^\mu$ with data prescribed on
$\mathcal{E}$.  This problem is \emph{ill posed} ---i.e. it is not
possible to establish local existence and uniqueness by standard
methods--- and only admits a solution if the spacetime is taken to be
analytic. Thus, under the hypothesis of analyticity it is possible to
show that $\eta^\mu$ extends to a Killing vector on the domain of
outer communication ---see \cite{HawEll73,Chr97}. This type of result
is known in the literature as a \emph{rigidity theorem} for it
shows that a certain assumption on the symmetry of the spacetime
---stationarity--- together with some further technical assumptions
imply further symmetries ---axial symmetry. The rigidity of black
holes has been recently revisited with the particular aim of
extending it from the $3+1$ dimensional to the $n+1$ setting
---cf. Isenberg's IHP talk \cite{IHP} and also \cite{HolIshWal06}.

The above argument for the \emph{uniqueness} of the Kerr black hole
has been criticised on the grounds that the analyticity assumption is
overly restrictive: general spacetimes are at most smooth, and it is
not obvious why a smooth spacetime will become analytic after it has
emitted gravitational radiation. In order to get around with this
problem an alternative approach to uniqueness has been given in
\cite{IonKla07a} ---see also Klainerman's IHP talk \cite{IHP}. The
idea behind this approach is that although the boundary value problem
for wave-like equations on $\mathcal{E}$ is ill posed, if one knows
that solutions do exists, then it is possible to show uniqueness
\cite{IonKla07b}. In \cite{Mar99,Mar00} a certain tensor ---the
\emph{Mars-Simon tensor}--- whose vanishing implies that the vacuum
spacetime under consideration is locally isometric to the Kerr
solution has been discussed. This tensor requires, for its definition
the existence of a stationary Killing vector. In \cite{IonKla07a} it
has been shown that the Mars-Simon tensor obeys a wave equation and
that it vanishes on $\mathcal{E}$ under a further technical assumption
---which does not require analyticity.  In general, one does not know
whether a solution to the boundary problem for the wave equation
satisfied by the Mars-Simon tensor with vanishing data on
$\mathcal{E}$ exists. However, if a solution exists, then it is
possible to argue that the Mars-Simon tensor vanishes not only in a
neighbourhood of the horizon but on the whole domain of outer
communication ---this is equivalent to show uniqueness of the ill
posed boundary value problem.  It is important to stress that this is
a global result and not only local to the horizon, like the one based
on the hypothesis of analyticity. Note further, that this proof
bypasses completely the use of rigidity results, and renders the
existence of axial symmetry as part of the main result.

\subsection{Quasi-local black holes}
\label{s:quasi_bh} 

In spite of the success and strength of the traditional formalisation
of the black hole notion in the asymptotic flatness setting
---cf. section \ref{s:BHs}--- black hole characterisation
in terms of elements not involving global
aspects of the spacetime has attracted significant efforts.
This area of research is particularly well suited
for a fruitful collaboration between mathematical and numerical
relativists.

\subsubsection{Motivations for quasi-local Black Holes.}

Quasi-local characterisations of black holes have been approached from
different communities addressing distinct
problems. Here, we highlight geometric motivations from the
mathematical community, on the one hand, and ``practical'' needs for
numerical relativists, on the other hand. Regarding the former, the
difficulties discussed in section \ref{s:BHs} related to the
understanding of the full implications of asymptotic flatness, and
illustrated in the absence of known examples of non-stationary
``strict'' black holes spacetimes, depicts a state of affairs which is
non-satisfactory at a conceptual level.  More dramatic are the issues
raised from a numerical perspective, where the required
global asymptotic information is simply not accesible during the 
evolution. Quasi-local
characterisations of black holes are needed both for ``technical''
reasons ---e.g. need of tracking the 
black holes along the evolution--- 
and for physical interpretative reasons 
---e.g. the need to disentangle information
about ``individual black holes'' in an interacting scenario.

Other motivations have played a crucial role in the historical
development of quasi-local notions of black holes: i) characterisation
of black holes in cosmological settings where asymptotic flatness
notions do not apply; ii) extension of black hole thermodynamical
results to situations where global notions are not under control; iii)
studies of black holes in the context of quantum gravity, such as the
microscopic evaluation of black hole entropy; and iv) conceptual
compliance with the consensus's notion of ``black hole'' in the
astrophysical community which is completely foreign to global issues.
In a general sense, it is fair to say that numerical and thermodynamical
issues have been the main drivers in the development of quasi-local
black holes as a line of research on its own.

\subsubsection{Seminal concepts and objects: trapped surfaces.}
\label{s:trapped_surfaces}

As discussed in section \ref{s:BHs}, the idea of a black hole as a
region of no escape can be formalised in terms of the set points not being
able of sending a signal to infinity ---this, as long as we have a
sensible notion of infinity. Local convergence of all the light rays
emitted from given sets of points provides an alternative approach to
encapture the notion of \emph{no-escape}.  This second approach does
not resort to a notion of infinity and provides the rationale for the
quasi-local approaches to black holes to be discussed in this section.

The seminal idea behind this characterisation is the concept of {\em
  trapped surface}. This notion has played a fundamental role in the
context of the singularity theorems \cite{Pen65b,HawPen70}. 
Given a surface
$\mathcal{S}$ which is spacelike orientable and closed ---i.e. compact
and without boundary--- one can unambiguously define at each point $p\in
{\cal S}$ two null directions $\ell^\mu$ and $k^\mu$, which span the
normal plane at $p$. That is, vector fields $\ell^\mu$ and $k^\mu$ span
the normal bundle $T^\perp{\cal S}$. The idea behind the notion of a
trapped surface ${\cal S}$ is that either all light rays emitted in
the normal null directions do locally converge or, at least, light
rays which would be naturally expected to expand, do actually
converge.  In the first case, the trapped character is an intrinsic
property of the surface ${\cal S}$, whereas in the second case some
extra structure is needed to make sense of the intuition of
\emph{naturally expanding} light rays. Let us define the expansions
$\theta^{(\ell)}$ and $\theta^{(k)}$, respectively associated with the
null congruences $\ell^\mu$ and $k^\mu$, as 
\bea
\label{expansions}
\theta^{(\ell)}= q^{\mu\nu}\nabla_\mu \ell_\nu \ \ , \ \ 
\theta^{(k)}= q^{\mu\nu}\nabla_\mu k_\nu \ \ ,
\eea
where $q_{\mu\nu}$ is the 
spacelike metric induced on ${\cal S}$ from the ambient spacetime metric.
Penrose \cite{Pen65b} presents an intrinsic notion
of trapped surface irrespective of any embedding.
A closed surface ${\cal S}$ is a {\em Trapped Surface} (TS) if and only if 
\bea
\label{e:TS}
\theta^{(\ell)}<0 \ \ , \ \ \theta^{(k)}<0 \ \ .
\eea
More generally, in order to cover Cosmological settings, a TS
is defined as satisfying $\theta^{(\ell)}\theta^{(k)}>0$, but 
we only discuss here the cases related to black holes.
The limiting case in which 
one the expansions ---say  $\theta^{(\ell)}$--- vanishes,
characterises ${\cal S}$ as a {\em Marginally Trapped Surface} (MTS):
\bea
\label{e:MTS}
\theta^{(\ell)}=0 \ \ , \ \ \theta^{(k)}<0 \ \ .
\eea
In the context of isolated systems it is natural to consider
surfaces ${\cal S}$ embedded in an asymptotically flat (Euclidean)
spacelike hypersurface $\Sigma$. In this case, we can naturally 
define the {\em outer} null direction, say $\ell^\mu$,
 as the one ``pointing''
towards infinity. Following Hawking \cite{HawEll73} we say that
a surface ${\cal S}$ is an {\em Outer Trapped Surface} (OTS) if and only if: 
\bea
\label{e:OTS}
\theta^{(\ell)}<0 \ \ .
\eea
The vanishing limiting case defines a {\em Marginally Outer Trapped Surface} (MOTS):
\bea
\label{e:MOTS}
\theta^{(\ell)}=0 \ \ .  
\eea 
Dropping the global condition this last
condition characterises the {\em marginal surfaces} defined by Hayward
\cite{Hay04}. In order simplify the discussion, we will refer
generically to (\ref{e:MOTS}) as the MOTS condition, and the context
will make it clear if global conditions are being taken into account
or not ---this is, in fact, the common practise in the literature,
e.g. in \cite{AndMarSim05}.  These are not the only attempts to
classify and characterise the idea that a surface is trapped. See in
this sense \cite{MarSen03,Senov07} for an alternative nomenclature,
\cite{HarWil73} for the introduction of {\em average trapped} surfaces
---cf. in this sense also \cite{Mal91}---
and \cite{Senov07b} for a geometric formulation of the {\em hoop
  conjecture} in terms of the notion of {\em trapped circle}.

Trapped surfaces open an avenue to control quasi-locally the notion of
black hole.  Given a spacelike hypersurface $\Sigma$, the {\em
  trapped region} ${\cal T}_\Sigma$ is defined in \cite{HawEll73} as the
set of points $p\in \Sigma$ belonging to some OTS, ${\cal S}\subset
\Sigma$.  The {\em Apparent Horizon} (AH) is defined as the outer
boundary of the trapped region ${\cal T}_\Sigma$. It attempts to
encapture quasi-locally the notion of black hole horizon at the
hypersurface $\Sigma$.  A crucial result is the characterisation,
under the appropriate regularity conditions, of the AH as the
outermost MOTS \cite{HawEll73}.  Even though these quasi-local notions
were originally developed in the mathematical studies of black holes,
particularly in the context of the gravitational collapse and
singularity theorems, they were very early employed in numerical
implementations.  

\paragraph{First numerical implementations using geometric quasi-local 
black hole ideas.}

Using their characterisation as MOTS, AHs have played a key role in
the relation between mathematical and numerical relativists. Firstly,
for  numerical relativists, condition (\ref{e:MOTS}) represents a
tentative                                                                                                                                              characterisation of ``black holes'' that can be efficiently calculated
in terms of data on a compact region of a 3-dimensional slice.
Important in this context, is the development of AH-finders, i.e.
algorithms to locate AHs in spatial slices by solving equation
(\ref{e:MOTS}) ---see \cite{Tho07} for a review and \cite{LinNov07} for 
a recent work; cf. also \cite{Metzg04} for related work with a strong geometric input.  
Moreover, under
appropriate energy conditions and assuming the weak Cosmic Censorship
conjecture, AHs are geometrically defined surfaces that are guaranteed
to lie inside the event horizon ---see e.g. \cite{HawEll73,Cla00}---
and therefore are causally disconnected from the rest of the
spacetime.  Although the extrapolation of this last feature from the
{\em continuum} description to the {\em discretised} level is not
straightforward ---see e.g. \cite{BroSarSch07}--- this geometric idea
has acted as a positive criterion in the development of black hole
evolution codes ---see section \ref{s:moving_punctures} for a brief
discussion of the different manners of exploiting this idea.

Together with the numerical control of the black hole during the
evolution, AHs were early applied as inner boundary conditions in the
construction of initial data for black hole spacetimes ---following an
idea proposed by Unruh.  In this case, condition (\ref{e:MOTS}) is
used to complete the elliptic system defined by the constraints
(\ref{Hamiltonian})-(\ref{Momentum}) when an interior sphere has been
removed ---cf. section \ref{s:moving_punctures} for further comments
on this {\em excision} approach.  This idea were first numerically
implemented by Thornburg \cite{Tho87}, later becoming a standard
technique. This problem has also received attention from the
mathematical community \cite{Dai04a,Max04b,Smith07} but unfortunately the
interaction between the communities seems to be scarce.  More
recently, the idea of AHs inner boundaries has also been proposed in
the evolution context by Eardley \cite{Eard97}. The difficulties in
implementing these ideas exemplify the needs of interaction between
geometric and numerical communities. See also \cite{Israe86a,Israe86b} for 
an alternative geometric solution to the inner excision problem 
in evolutions using, instead of AHs, 
locally area-preserving evolutions of a given initial trapped surface.

\paragraph{The need of a spacetime point of view.} In spite of their
conceptual and practical interest, AHs do not provide a spacetime
characterisation of black holes. The AH notion depends on
the spacelike slice $\Sigma$ and this feature limits its applicability.
The latter has been illustrated in examples given by Wald \& Iyer
\cite{WalIye91} of Schwarzschild slicings where Cauchy hypersurfaces
come arbitrarily close to the singularity but do not contain any OTS.
In particular, if we understand the evolution of an AH, say ${\cal S}_0$,
as the ``hypersurface'' formed by piling up the AHs 
---denoted by ${\cal S}_t$---
found in the 3-slices $\Sigma_t$ used in a
Cauchy evolution ---that is, $\Sigma_t\supset {\cal S}_t$---
the resulting AH-worldtube $\bigcup_{t} {\cal S}_t$
explicitly depends on the chosen 3+1 slicing. Furthermore the
geometric, dynamical and thermodynamical properties of this
AH-worldtube are not under control ---in particular non-continuous
jumps can occur.  This contrasts with event horizons, that are smooth
null hypersurfaces under the appropriate conditions \cite{HawEll73}.

An intrinsic spacetime formulation can be obtained by defining the
trapped region without any reference to a particular slice $\Sigma$.
In this context, the {\em trapped region} ${\cal T}$ of a spacetime
${\cal M}$ is the set of points $p\in {\cal M}$ belonging to some TS,
${\cal S}\subset {\cal M}$.  The outer boundary of this region is
sometimes also referred in the literature as the AH, but following
\cite{Hay94} we will denote it as the {\em trapping boundary}.  This
offers an intrinsic quasi-local notion for the horizon
where no reference to asymptotic
quantities is needed.  A natural question to address consists in
clarifying its relation with the event horizon ---if the latter can be
defined.  As in the AH case, assuming Cosmic Censorship, the {\em
  trapping boundary} do not extend beyond the event horizon. Finding
out if it actually extends up to it, or if it remains strictly in the
interior, is an open issue which can be of relevance in the
understanding of Cosmic Censorship Conjecture and in which
mathematical and numerical collaboration can prove to be very useful
---cf. ideas and results in
\cite{Hay94,Eard97,KriHay97,SchKri05,BenDo06}.  However, in spite of
its conceptual interest this notion also entails ``practical''
difficulties. Namely, the trapping boundary is a complicate object to
locate, in particular in a numerical evolution.  And secondly,
regularity issues may arise on this horizon \cite{Hay94}.

\subsubsection{Evolution of marginally trapped surfaces: quasi-local
  black holes.} The need of an operational quasi-local notion of black
hole horizon in an evolution has led to the formalisation of the idea
of world-tubes of AHs or, more generally, world-tubes of MOTS
---understood here as {\em marginal surfaces} in the sense of
\cite{Hay94}. This is done at the price of loosing uniqueness.

First steps where taken by H\'a\'\j i\v{c}ek in
\cite{Haj73,Haj74a,Haj74b} were he introduced {\em perfect horizons}
as non-expanding null hypersurfaces devised to model stationary black
hole horizons. These totally geodesic null hypersurfaces \cite{Haj73}
were the first examples of world-tubes of AHs where not only the
expansion $\theta^{(\ell)}$ vanishes but also the shear of the null
congruence does.  A more systematic and general approach was initiated
by Hayward in \cite{Hay94}, where he introduced {\em future outer
  trapping horizons} (FOTH) to model quasi-local black holes in
generic situations.  A parallel line of research in quasi-local black
holes as AH world-tubes has been developed by Ashtekar and
collaborators. This has lead to \emph{isolated horizons} (IH)
\cite{AshBeeFai99,AshCorKra99,Ash_al00,AshBeeFai00,AshFaiKri00,AshBeeLew01,AshBeeLew02,AshEngPaw04}
and, together with Krishnan, to \emph{dynamical horizons} (DH) 
\cite{AshKri02,AshKri03}.  FOTHs, on the
one hand, and IHs together with DHs, on the other hand, offer
complementary approaches to the quasi-local black hole problem.
Whereas FOTHs have the virtue of providing a general single
characterisation which encompasses both stationary and dynamical
situations, the asset of IHs and DHs lies in their adaptation to the
specific and rather different geometric structures associated with
null (stationary) and spacelike (dynamical) AH-worldtubes.  Both
approaches underline different aspects of the problem and conceptual
compatibility is guaranteed since equivalence have been shown
in {\em generic} conditions 
 \cite{AshGal05,AndMarSim05,BooFai07} ---see also \cite{Kor06}
and later in this section.  Finally, Booth
\& Fairhurst \cite{BooFai04,BooFai07,KavBoo06} have introduced the
notion of {\em slowly evolving dynamical horizons} to address the
physical issues happening in the transition from the stationary to the
dynamical regimes.  General review articles on these quasi-local
approaches can be found in \cite{AshKri04,Boo05,GouJar06,Kri07}. We briefly
review the ideas contained in them.

\paragraph{Trapping horizons.} A {\em trapping horizon} is (the
closure of) a hypersurface ${\cal H}$ foliated by MOTS \cite{Hay94}
---this notion has applicability beyond the context of black holes,
in Cosmological contexts.
Here the MOTS condition $\theta^{(\ell)}=0$ must be read in its local
sense as a {\em marginal surface}. White and black hole situations are
distinguished by the sign of $\theta^{(k)}$, whereas the (local) outer-
or innermost character is captured by the sign of
$\delta_{k}\theta^{(\ell)}$ ---where $\delta_k$ denotes formally a {\em
  variation} in the direction of the vector $k^\mu$. Trapping horizons
are originally formulated in a double-null foliation of spacetime
\cite{Hay94}, where $\delta$ becomes a Lie derivative. In this sense,
trapping horizons are:
\begin{itemize}
\item[ i)] {\em Future} if $\theta^{(k)}<0$, or 
{\em past} if $\theta^{(k)}>0$.
\item[ ii)] {\em Outer} if 
$\delta_{k}\theta^{(\ell)}<0$, or {\em inner} if $\delta_{k}\theta^{(\ell)}>0$.
\end{itemize}
Note that the extremal case $\delta_{k}\theta^{(\ell)}=0$ is not
treated ---see section \ref{s:geom_ineq}.  Inner trapping horizons are
of interest in Cosmological scenarios.  But in quasi-local black hole
horizon settings, {\em inner} light are expected to converge
---that is, $\theta^{(k)}<0$--- whereas {\em outer} light rays should converge
just inside the black hole and diverge just outside
---that is, $\delta_{k}\theta^{(\ell)}<0$. Therefore, the adequate
characterisation is in terms of {\em future outer trapping horizons}, 
hypersurfaces ${\cal H}$ foliated by MTSs $\mathcal{S}_t$ 
on which the {\it stability}
condition $\delta_{k}\theta^{(\ell)}<0$ holds.
 
We comment on some fundamental features of FOTHs. Under the dominant
energy condition, the MTS slices $\mathcal{S}_t$ of a FOTH have 
spherical topology ---the
{\em topology law}.  Under the null energy condition: i) the vector
tangent to the FOTH and normal to the MOTS slices can be either
spacelike or null ---the point-wise {\em signature law}--- the latter
case happening if both the shear $\sigma^{(\ell)}$ and $T(\ell,\ell)$
vanish, and ii) the area of the MTS slices remains constant if ${\cal
  H}$ is null and grows otherwise ({\em area law}).  Property i) follows
from $\delta_{k}\theta^{(\ell)}<0$ and formalises the idea of 
AH-worldtubes as being stationary (null) if nothing falls into the horizon and
superluminal (spatial) otherwise, whereas property ii) is the
quasi-local counterpart of Hawking's area theorem \cite{Hawki71,Hawki72}.

A main application of FOTHs has been the derivation of balance laws
for physical quantities on the horizon, and its subsequent application
to black hole dynamics ---see Hayward's IHP talk \cite{IHP} and below.  
Regarding the conceptual
characterisation of quasi-local black holes, Hayward has also shown
that under a regularity assumption ---the existence of foliation of
${\cal H}$ by {\em limit sections} \cite{Hay94}--- the trapping
boundary is actually a trapping horizon, in fact the outermost.

\paragraph{Isolated horizons and dynamical horizons.} 
The important geometric differences between null and spatial hypersurfaces
suggest the use of different strategies for addressing specific issues 
in the stationary and dynamical regime. 

The framework of isolated horizons ---see \cite{AshKri04} for a very
comprehensive account--- provides a hierarchy of geometric structures
built on a null hypersurface, and devised in order to capture
different stationarity levels of a black hole placed in an otherwise
dynamical environment.

{\em a)} {\em Non-expanding horizons} (NEH) provide the minimal notion
of stationarity. They are defined as null hypersurfaces ${\cal H}$ of
$\Sphere^2\times \mathbb{R}$ topology where: i) the null generator
expansion $\theta^{(\ell)}$ vanishes, and ii) the vector $-{T^\mu}_\rho
\ell^\rho$ is future directed and causal. Using the (null)
Raychaudhuri equation, a NEH ${\cal H}$ is a hypersurface foliated by
MOTS and characterised by the vanishing of both $\sigma^{(\ell)}=0$
and $T(\ell,\ell)$.  The geometry of a NEH is given in terms of
$(q_{\mu\nu}, \hat{\nabla})$, where $q_{\mu\nu}$ is the spatial metric
induced on any compact slice ${\cal S}$ and $\hat{\nabla}$ is a
connection uniquely induced from the ambient spacetime connection $\nabla$.
The NEH condition expresses the evolution invariance of the intrinsic
geometry: ${\cal L}_\ell q_{\mu\nu}=0$, and essentially coincides with
the notions of {\em perfect horizons} and {\em null trapping horizons}.

{\em b) Weakly isolated horizons} (WIH) are NEHs
with some additional structure needed for the analysis of equilibrium black hole
dynamics: an equivalence class of null normals $[\ell]$
for which the {\em surface gravities} $\kappa_{\ell}$, defined from 
$\ell^\rho \nabla_\rho \ell^\mu = \kappa_{\ell} \ell^\mu$, are constant on
${\cal H}$ ---the {\em zeroth law}. A Hamiltonian analysis
of the symplectic space of solutions of the EFE containing a WIH
leads to a (Gibbs-like) quasi-local first law of black hole 
thermodynamics, providing
in particular a quasi-local expression for the mass of a 
black hole with a stationary horizon.

{\em c) Isolated horizons} (IH) are NEHs where the {\em extrinsic
  geometry} $\hat{\nabla}$ is also invariant long the evolution:
$[{\cal L}_\ell, \hat{\nabla}]=0$.  They represent the strongest
quasi-local notion of stationarity and crucially allow the definition
of mass and angular momentum ---in the axisymmetric case--- multipoles
\cite{AshEngPaw04}.  The richness of the Isolated Horizon framework
has led applications in very diverse areas of black hole physics, such
as the microscopic evaluation of entropy in the context of Loop
Quantum Gravity \cite{AshBaeCor98,Ash99,AshBaeKra00}, or the study of
properties of hairy black holes and solitonic solutions in theories involving the coupling
with additional fields ---e.g. Yang-Mills, dilaton, Higgs, Proca and Skyrme
fields \cite{AshKri04}.

Regarding the dynamical case, Ashtekar \& Krishnan introduced in
\cite{AshKri02,AshKri03} the notion of {\em dynamical horizons} as
spatial hypersurfaces ${\cal H}$ that can be foliated by MTSs 
---surfaces $\mathcal{S}_t$ where (\ref{e:MTS}) holds. 
Note the difference with the FOTH definition, where the
local outermost condition $\delta_k\theta^{(\ell)}<0$ has been
substituted by the condition on the spatial character of ${\cal H}$.
FOTHS and DHs are equivalent in generic circumstances \cite{AshGal05,BooFai07}.
A FOTH for which $\delta_\ell\theta^{(\ell)}\neq 0$ for at least one
point on each MTS section, is a DH. For the converse, a {\em genericity}
condition ---namely $\delta_\ell\theta^{(\ell)}\neq 0$ everywhere on 
${\cal H}$--- and the null energy condition are required. Examples
of DHs not satisfying the genericity condition, and thus failing to be 
FOTHs, are presented in \cite{Senov03} ---see also comments in
section \ref{s:geom_ineq} in the context of extremal DHs. 
In contrast with the double null formalism of
FOTHs, DHs are formulated in a 3+1 framework. Therefore they are specially
suited for their use in numerical relativity. Although
black hole mechanics relations can and have been derived in this
setting, a strong emphasis is put in the application of the associated
balance equations to the analysis and test of numerical simulations in
the strong field regime.

\paragraph{Quasi-local Black Hole thermodynamics.} The extension of
classical black hole thermodynamical results \cite{BarCarHaw73} from
stationary spacetimes to more generic situations has been one of the
main motivations for the development of quasi-local black holes ---see
Hayward's IHP talk \cite{IHP}.  Different approaches and definitions
of the physical parameters have led to distinct existing versions of
the generalised thermodynamics laws, whose detailed review goes beyond
the scope of this article. We comment in general terms: i) {\em Zeroth
  law}: constancy of the surface gravity, defined as the non-affinity
parameter associated with the generating null vector in the context of
isolated horizons, it is shown to characterise a WIH \cite{AshKri04}
---and therefore holds on an IH.  A generalised zeroth law in terms of
an inequality bounding the mean of the ({\em trapping}) gravity is
proposed in \cite{Hay94}, where the constancy of $\kappa$ is
characterised by the saturation of the inequality ---see also \cite{NieYoo07}
for a general discussion on the surface gravity in the quasi-local
context. ii) The {\em second
  law}, non-decrease of the area, follows from the spatial (or null
character) of the MOTS-worldtubes, together with the condition
$\theta^{(k)}<0$. Therefore, the second law is built into the
definition of quasi-local black holes.  iii) The discussion of the
first law, as a Gibbs-like expression relating the variation of the
energy, area and angular momentum of the horizon, is more problematic
due to the ambiguities in defining quasi-local physical parameters
---cf. section \ref{s:quasi-local_param}.
In the spherically symmetric and axisymmetric IH cases, this has been
successfully addressed ---see \cite{AshFaiKri00,AshBeeLew01}--- and
has lead to a first law in terms of variations in the space of
physical states ---Gibbs' version.  However, the attempt of deriving a
first law relating time variations of physical quantities (Clausius'
version) has led to different versions according to the distinct
choices of evolution vector or quasi-local energy. For these reasons, the status
of such a version of the first law in the dynamical regime is unclear.
In this context, the ensemble of {\em balance} or {\em
  conservation laws} obtained from the restriction of different
components of the Einstein equation on the MOTS-worldtubes ${\cal H}$
acquire a particular relevance.
We highlight the balance equation relating the variation
of the area and angular momentum to the flux of {\em energy}
\cite{Hay94,AshKri02,AshKri03,Hay04,Hay04b,BooFai04,BooFai05} ---see
\cite{BooFai07} for a discussion of the comparison between some of
them.  A set of evolution equations for the area
\cite{Hay94,AshKri03,AshKri04,BooFai04,GouJar06b}, angular momentum
\cite{AshKri02,AshKri02,BooFai04,BooFai05,Gou05,Hay06} or the charge
\cite{Hay06b} have also been derived.  In particular, references
\cite{Gou05,GouJar06,GouJar06b} discuss a quasi-local version of the
membrane paradigm \cite{Damou79,Damou82,PriTho86,ThoPriMac86} based on a
hydrodynamical analogy between horizons and viscous fluids --- whose
full analysis will plausibly involve the use of concepts and
techniques from PDE theory of hyperbolic systems.

We must emphasise that some of the ambiguities present in the fully
dynamical regime can be controlled when considering slight deviations
from equilibrium, in the setting of Booth \& Fairhurst's {\em slowly
  evolving dynamical horizons}.  Finally, regarding the impossibility
of reaching extremal black holes ({\em third law}), a discussion of
extremality in this quasi-local context has been recently developed in
\cite{BooFai07b,Boo07}.

\subsubsection{Numerical implementations of dynamical trapping horizons.}
The implementation of ideas from dynamical trapping horizons offers an
example of a fruitful relation between Geometry and Numerics. The
application of some prescriptions derived from the framework of dynamical
trapping horizon have been used to extract physical information about
numerically constructed black holes. Conversely, numerical experiments
have provided key insights into the geometric structure of
MOTS-worldtubes.

\paragraph{DHs and ``a posteriori'' analyses.} The main application of DHs in
numerics is the physical and geometric analysis of MOTS-worltubes located
by an AH-finder along a numerical evolution.  Krishnan's IHP presentation
\cite{IHP} provides an excellent account of this kind of analysis.
Regarding the extraction of physics, DHs have been used to extract
individual black hole masses and spins in black hole spacetimes
\cite{Bai_al04,SchKriBey06,CamLouZlo07,KriLouZlo07}.  More
specifically, deformation of black holes can be analysed by computing
mass and angular momentum DH multipoles \cite{SchKriBey06}. 
DHs can be employed to characterise the rate of approach to
Kerr \cite{SchKriBey06} or to study spin-orbit effects in binary black
hole simulations \cite{CamLouZlo07}. Heuristic approaches can be formulated
in this setting for determining a quasi-local linear momentum of a black hole,
very relevant in the astrophysical study of recoil velocities after mergers \cite{KriLouZlo07}.
 From a geometric point 
of view, the results in \cite{SchKriBey06}
support the picture that AHs jumps occurring in black hole evolutions
correspond to a situation in which the involved DHs are actually
connected by a world-tube of MOTS which violates some of the FOTH
conditions. The resulting  single MOTS-worldtube presents timelike or 
signature-mixed sections where topology change is
possible. This picture is also supported by analytic results
in \cite{BooBriGon05} on the Tolman-Bondi collapse ---see also \cite{NieVis06}.  
A prediction of
this picture is that in binary black hole evolutions modelled by two
MOTS-worldtubes ${\cal H}_1$ and ${\cal H}_2$, right after the
formation of the outer common horizon ${\cal S}_o$, the latter should
{\em bifurcate} into two MOTS-worldtubes: an exterior spacelike one
${\cal H}_{\mathrm{outer}}$ of growing area, and an interior one
${\cal H}_{\mathrm{inner}}$ of mixed signature evolving into a
timelike MOTS-worltube. This is actually seen in numerical experiments
\cite{SchKriBey06}.  Under evolution, this ${\cal H}_{\mathrm{inner}}$
should annihilate with the {\em ghost} horizons ${\cal H}_1$ and
${\cal H}_2$, but the details of this process are unknown. In this context,
the assessment of the recent results in \cite{SziPolRez07} showing the
overlap of the inner horizons ${\cal H}_1$ and ${\cal H}_2$ is of
special relevance.

\paragraph{DHs as ``a priori'' ingredient.}
An alternative application of quasi-local black holes to numerics is
their use as a constitutive element in the PDE evolution system.  This
is exemplified by the use of NEHs to prescribe inner boundary
conditions in the construction of initial data describing spacetimes
containing black holes in instantaneous equilibrium ---see e.g.
\cite{CooPfe04,Ans05,CauCooGri06,JarGouMen04,DaiJarKri04,GouJar06,JarAnsLim07}.
A scheme for using MOTS ---and not MTS--- as inner boundary
conditions in the context of constrained evolution formalisms has been
presented in \cite{JarGouCor07} and at Gourgoulhon \& Jaramillo IHP talk
\cite{IHP}.  This essentially recasts Eardley
programme \cite{Eard97} in the dynamical trapping horizon framework. Its
feasibility must still be assessed.

\subsubsection{Geometric Analysis.}

A fundamental shift in the research on quasi-local horizons has
occurred with the application of tools of geometric analysis to the
understanding of dynamical horizons properties.  Besides the
continuing efforts to characterise DH physical parameters \cite{Kor07}
and derive appropriate {\em conservation equations} for them, some
effort has been put into the use of maximum principles and related
notions to the study of the properties of elliptic equations defined
on horizon sections.

The methods of geometric analysis has been put into work to show that,
under appropriate conditions, FOTHs can be fully partitioned into NEH
and DH sections ---see \cite{AndMarSim05,BooFai07}. This result rules
out the possibility of finding MOTS sections in FOTHs that are
partially null and partially spacelike ---a possibility which was not
discarded in early works. This means that the transition from
equilibrium to the dynamical regime happens {\em all at once}. This
shows the complete equivalence, in generic circumstances, 
between the two main approaches to
quasi-local black holes ---at least in what regards equilibrium and
the dynamical stages of the quasi-local horizons.

Two results deserve special mention. Firstly, the {\em foliation
  uniqueness theorem} for dynamical horizons by Ashtekar \& Galloway
\cite{AshGal05}: given a dynamical horizon ${\cal H}$, its foliation
by MTSs is unique. Secondly, a {\em local existence theorem} by
Andersson, Mars and Simon \cite{AndMarSim05} stating that, given a 3+1
slicing $(\Sigma_t)$ and an initial MTS ${\cal S}_o\subset \Sigma_o$
satisfying a stability condition closely related to the {\em outer}
condition of FOTHs, there exists a unique worldtube foliated by MTSs
${\cal S}_t$, such that ${\cal S}_t\subset \Sigma_t$ ---at least as
long as the stability condition is satisfied.

Some more recent developments in the geometric study of dynamical
trapping horizons include the refinement of previous work
\cite{AndMarSim05} on the stability and existence analysis of
MOTS-worldtubes by Andersson, Mars and Simon in \cite{AndMarSim07};
the derivation of estimates for the curvature in MOTS with 
application on DHs \cite{AndMet05}; studies by Bartnik \&
Isenberg \cite{BarIse06} of necessary and sufficient conditions for
the existence of DHs in spherically symmetric spacetimes; the analysis
of the asymptotics of MOTS-worldtubes in spherically symmetric
spacetimes \cite{Wil07}; the formulation of a conjecture about
the {\em peeling behaviour} of DHs \cite{Senov07b} in the context 
of the geometric discussion of the hoop conjecture; or
the study of area
estimates for outermost MOTS and the characterisation of the boundary
of the trapped region \cite{AndMet07} ---see also \cite{Metzg07,CarMar07}.
Finally, we mention an approach towards a general proof of Penrose
inequality {\em \`a la Huisken-Ilmanen} \cite{HuiIlm01} using a
spacetime generalisation of the inverse mean curvature flows in terms
of uniformly expanding flows \cite{BraHayMar06}. These developments are
very close in spirit to some of the possible applications of dynamical
trapping horizons \cite{And06}.

\subsubsection{General perspective.}

The recent numerical and geometric insights have enriched the research
in quasi-local horizons, which was previously focused mainly on black
hole thermodynamical aspects. From the DH existence and foliation
uniqueness theorems \cite{AshGal05,AndMarSim05} it follows that the
question about {\em the evolution} of a given initial MTS, ${\cal
  S}_o$, is not well-defined in generic situations. The initial ${\cal
  S}_o$ can evolve into different DHs depending on the chosen
foliation ---i.e. on the chosen lapse. Although this can be completely
harmless in most ``practical'' numerical situations, it is
conceptually important and was not sufficiently stressed in the
numerical community. In numerical evolutions, black holes
characterised by MOTSs world-tubes are treated as other {\em standard}
compact objects. In particular, this assumes a well-defined unique
evolution independent of the {\em observer}. The clarification of this
point permits to turn the argument around, opening the possibility of
setting a {\em preferred} choice of lapse function in terms of a
geometrically single out DH ---first attempts in this direction are
discussed in \cite{GouJar06b}.  In a different line of work, numerical
results by Schnetter and Krishnan ---see also \cite{BooBriGon05}---
have shed light on the global structure of tubes of MOTS, and on the
changes of signature along the evolution and the geometric criteria to
control them. Other issues, such as the study of generic dynamical
trapping horizon asymptotics towards the event horizon, and its
possible application to fundamental problems such as Kerr stability
\cite{And06} or Penrose inequality will probably require the
combination of analytic, geometric and numerical skills.

\subsection{Geometric inequalities involving black hole horizons}
\label{s:geom_ineq}
As in the case of other physical theories, geometric inequalities in
GR are often the reflect of a fundamental underlying physical
principle.  A prime example of this is the {\em positivity of mass} in
GR \cite{SchYau79,SchYau81,Wit81}. The fundamental nature of this
result is manifest from its role in many crucial developments in GR.
Furthermore, its failure would put the physical consistency of the
theory under question.

In this section we will briefly comment on some geometric inequalities
in the context of black hole spacetimes which, in particular,
constraint the gravitational collapse process.  Our present
understanding of the gravitational collapse is based on a chain of
results and conjectures. First, the singularity theorems
\cite{Pen65b,Hawki67,HawPen70,HawEll73} guarantee that the appearance
of a trapped surface during the gravitational collapse leads to the
development of a singular spacetime. Second, the singularity should be
hidden behind a black hole event horizon so as to avoid a lack of
predictability ---this physically motivated hypothesis excluding the
formation of naked singularities was proposed by Penrose \cite{Pen69}
and is known as the {\em weak Cosmic Censorship Conjecture}. Third,
the black hole spacetime should reach a stationary state ---this assumption
is justified by the finiteness of the amount of radiation that an isolated
system can emit. And fourth, assuming that all fields have fallen into
the black hole after some finite time, the black hole uniqueness
theorems \cite{Heusl98} then guarantee that the spacetime settles down to a Kerr black
hole.  Thus, barring some technical
assumptions, if enough matter concentrates in a sufficiently compact
region, then the system evolves to a final Kerr provided {\em weak
  Cosmic Censorship Conjecture} holds and a final stationary state is
reached.

Using a chain of heuristic arguments based on the previous standard
picture ---the so-called \emph{establishment picture}--- Penrose
proposed \cite{Pen73} a lower bound for the total (ADM) mass of a
black hole spacetime in terms of the square root of the {\em area of
  the black hole}.  This {\em Penrose inequality} provides a lower
bound for the black hole contribution to the total mass. It
conjectures, in particular, a significant strengthening of {\em
  positive mass theorems} in the black hole context.  In its first
version ---that can be referred to as {\em global}--- this {\em
  Penrose inequality} conjectures that the area $A_{\cal E}$ of any
section of the event horizon ${\cal E}$ satisfies
$M_{_{\mathrm{ADM}}} \geq \sqrt{A_{\cal E}/16\pi}$.  Remarkably, it
can be formulated as a problem for initial data on a Cauchy surface
$\Sigma$, providing a version {\em local in time} of the Penrose
inequality. Given complete, asymptotically flat Cauchy data on
$\Sigma$ satisfying the {\em dominant energy condition}, the Penrose
inequality ---in the formulation of \cite{Hor84}--- conjectures
 \bea
\label{e:Penrose_ineq}
A_{\mathrm min} \leq 16 \pi M_{_{\mathrm{ADM}}}^2 \ \ , 
\eea 
where
$A_{\mathrm min}$ is the minimal area enclosing the apparent horizon
---cf. \cite{BenDo04} for an explicit construction illustrating the 
need of using the area of that minimal surface, rather than
the outermost MOTS in $\Sigma$.  There is a {\em rigidity}
side to the conjecture. Namely, that the equality is only attained in the
spherically symmetric (Schwarzschild) case.  The Penrose inequality
was proposed in an attempt to provide evidence of the violation of
weak Cosmic Censorship. Other inequalities involving minimal surfaces in
\cite{Gib72} ---employed to provide lower bounds for areas of event
horizon sections, as well as upper bounds for the efficiency in the
emission of gravitational radiation--- were also constructed with the aim
of providing evidence against Cosmic Censorship.  However, growing
evidence has accumulated in time on the {\em generic} validity of weak
Cosmic Censorship \cite{Wal97} and the effort has shifted to the
construction of a proof of Penrose inequality.  Although the latter
clearly does not imply the correctness of the standard gravitational
collapse picture, it would actually provide a strong support for it.
The spherically symmetric case has been proved in \cite{MalOMu94}. 
Huisken \& Ilmanen \cite{HuiIlm01} and Bray \cite{Bra01} have provided
independent proofs for the so-called {\em Riemannian} case ---where
$K_{ij}=0$--- but a general result has not yet been obtained. The
Penrose inequality has evolved into a problem in its own right,
becoming one of the important challenges in GR and Differential
Geometry. Its alternative name as the {\em isoperimetric inequality
  for black holes} \cite{Gib84,Gib97,GibHol06} underlines its
intrinsic geometric relevance and might ``lead to the importation into
black hole theory of further useful ideas and techniques from global
analysis'' \cite{Gib97} ---see e.g. \cite{BizMalOMu88,BizMalOMu89,Mal91,Mal92,MalOMu94b,GuvOMu97} for 
references related to 
isoperimetric inequalities in this context. Reviews discussing the original Penrose
argument, historical developments, main results and open questions can
be found in \cite{BraChr04,Mar07}. For early attempts 
to probe numerically the Riemannian case see \cite{KarMalSwi93,KarKocSwi94},
and for some more recent numerical studies 
see, for example, \cite{DaiLouTak02,KarMal04,Karko06,JarVasAns07}.

There exist some generalisations of the Penrose inequality which
involve linear momentum \cite{MalMarSim02}, charges inside the
apparent horizon \cite{GibHawHor83,Her98,WeiYam05} ---cf. also \cite{MalRos1998}
and related works \cite{MalOMu94,IriMalOMu95}--- or a cosmological
constant ---cf. \cite{ChrSim00,Gib99} for asymptotically anti-de
Sitter spacetimes.  Here we comment further on a particular sharpened
version involving the angular momentum $J$ in axially symmetric
spacetimes \cite{DaiLouTak02,Hawki72}:
 \bea
\label{e:Penrose_J}
A \leq 8\pi \left(M_{_{\mathrm{ADM}}}^2 + \sqrt{ M_{_{\mathrm{ADM}}}^4
    - J^2}\;\right ) \ \ .  
\eea 
A \emph{rigidity} property
---\emph{Dain's rigidity conjecture}--- has been explicitly formulated in
\cite{DaiLouTak02}: the equality holds if and only if the (exterior)
initial data corresponds to (exterior) Kerr data. Defining
$\epsilon_{_A} := A / \left[8\pi \left(M_{_{\mathrm{ADM}}}^2 + \sqrt{
      M_{_{\mathrm{ADM}}}^4 - J^2}\;\right )\right]$, inequality
(\ref{e:Penrose_J}) is expressed as $\epsilon_{_A} \leq 1$ and Dain's
conjecture, $\epsilon_{_A} = 1$, characterises Kerr data by the
evaluation of a single real number ---see
\cite{DaiLouTak02,JarVasAns07} for possible numerical applications. In
case of being true, this would strengthen current results on
characterisations of Kerr \cite{Mar99,Mar00} and Schwarzschild
\cite{GarVal07} involving the evaluation of tensor quantities. The
non-triviality of Dain's proposal can be appreciated in inequality
(37) of \cite{Mar07}. The latter would provide the counterpart for a variational
characterisation of Reissner-Nordstr\"om data, but has been
found to be false \cite{WeiYam05}.

A necessary condition for inequality (\ref{e:Penrose_J}) to make
sense, is the positivity of the quantity under the square root symbol.
This leads to the consideration of mass-angular momentum inequalities
in vacuum ---in presence of matter such an inequality is easily
violated. The data has to be subextremal.  In a series of articles
\cite{Dai06a,Dai06b,Dai06c,Dai06d,Dai07} Dain has proved the validity
of $ |J| \leq M_{_{\mathrm{ADM}}}^2$ for maximal, vacuum,
asymptotically flat, axisymmetric initial data. Equality is only
reached for extremal Kerr.  This theorem can be seen as a first step
in the study of the non-linear stability of Kerr. These results have
been discussed by Chru\'{s}ciel in his IHP talk, and further developed
in \cite{Chr07}.  In spite of its characterisation in terms of initial
data, these mass-angular momentum inequalities are spacetime
properties. There has also been an interest to study some quasi-local 
versions of these inequalities involving, in particular, the local
characterisation of extremality.  
In a first attempt, Ansorg and
Petroff \cite{AnsPet05,PetAns05,AnsPet06} have considered the
substitution of the ADM mass by the Komar mass $M_{_{\mathrm{Komar}}}$
evaluated on the black hole horizon of stationary and axially
symmetric spacetimes.  This has led to the numerical construction of
stationary configurations of a black hole surrounded by a matter torus
where the quotient $|J|/M_{_{\mathrm{Komar}}}^2$ could reach arbitrary
high values \cite{AnsPet05,PetAns05} or even $M_{_{\mathrm{Komar}}}$
could become negative \cite{AnsPet06}.  Instead, and in order to refer only to intrinsic
quantities on the horizon, one could consider using the irreducible
mass. This has led Petroff and Ansorg ---in Ansorg's IHP talk \cite{IHP}---
to conjecture an inequality for axisymmetric
stationary spacetimes only involving the area:
$8\pi|J| \leq A$.  This conjecture has been further developed
---including the charged case--- in \cite{AnsPfi07}, where use is made
of the Christodoulou-Ruffini mass and it is
analytically shown that {\em extremality} is characterised by the
saturation of the area-angular momentum-charge inequality.
In parallel, Booth \& Fairhurst \cite{BooFai07b,Boo07} have undertaken an
analysis of the local characterisation of extremality of black holes
based on the dynamical trapping horizon framework of section
\ref{s:quasi_bh}. After also considering the use of geometric
inequalities involving the area $A$ ---or alternatively, the vanishing
of a locally defined surface gravity--- they have opted for a
characterisation of extremality in terms of the absence of trapped
surfaces inside the apparent horizon. In terms of the {\em outer}/{\em
  inner} trapping horizon characterisation \cite{Hay94}, this reads
$\delta_k \theta^{(\ell)}=0$, and leads to the introduction of a
quasi-local parameter $e$ satisfying $e\leq 1$, such that extremality
corresponds to $e=1$. DHs satisfying the
genericity conditions \cite{AshGal05,BooFai07} referred to in section
\ref{s:quasi_bh}, are found to be subextremal in this sense. In fact, 
dropping these DH genericity conditions is equivalent to 
the extremal characterization by $\delta_k \theta^{(\ell)}=0$ ---precisely the feature
exploited in \cite{Senov03} to construct examples
of spacetimes containing DHs but without trapped surfaces.

Before concluding this section on geometric inequalities involving
black holes, we must briefly comment on the so-called {\em hoop
  conjecture}. It proposes that black hole horizons happen whenever
matter gets sufficiently compacted in {\em all spatial}
directions. Stated in an intentionally vague manner, ``black holes
with horizons form when, and only when, a mass $M$ gets compacted into
a region whose circumference $C$ in every direction satisfies $C
\lesssim 4\pi M$'' \cite{MisThoWhe, Thorn72}. In particular, the hoop
conjecture
offer an interesting example of interaction between geometry and
numerics ---see e.g. \cite{ChiNakNak94,ShaTeu91,Chiba99,ChiMae94,YosNamTom02}
for some numerical studies. Recently, a
reformulation of this conjecture as a genuine and mathematically sound
---see also \cite{Mal91}--- geometric inequality has been 
presented in \cite{Senov07b} ---see also
references therein for a review of the original conjecture.

Black hole geometric inequalities offer a link between conceptual
issues in GR such as Cosmic Censorship, positivity of mass or black
hole dynamics ---the latter, through the role of the area in the
second law.  Furthermore, its constraining role in the gravitational
collapse process is of potential interest in numerical constructions.

\subsection{Binary black holes}

The binary black hole problem has been the main challenge for the
numerical relativists in the last decades. The relevance of this
problem lies, on the one hand, on its conceptual richness ---this
two-body problem provides a probe into the strong field non-linear
regime of General Relativity--- and, on the other hand, on its
astrophysical interest ---stellar and supermassive black holes
constitute some of the main candidates for the detection of
gravitational waves signals by interferometric antennae. Here, we
emphasise the role of the binary black hole problem as a laboratory
for the development and test of new numerical, analytical and
geometrical ideas in GR.  The developments in the numerical evolutions
of binary black hole spacetimes through the inspiral \cite{BruTicJan04}, merger and
ringdown phases together with the extraction of gravitational
radiation \cite{Pre05,CamLouMar06,BakCenCho06} 
have constituted a
milestone. Since then, quite a large number of groups have succeeded
in developing binary black hole evolution codes. The last couple of
years have witnessed a rush in scientific activity. Given the volume
of the associated literature, we refer the reader to the review
\cite{Pre07}.  We limit here ourselves to discuss some points of
interest in the interaction between Geometry and Numerics.

\subsubsection{Helical Killing vectors and binary black hole initial data.}
\label{s:helical}
Gravitational radiation reaction drives relativistic systems into
inspiral motion, circularising the orbits very efficiently at least
for comparable mass systems \cite{Bla06}.  Therefore, for two bodies
sufficiently separated, it is natural to approximate the spiral orbits
by closed circular ones. This physical image is geometrically
en-captured by the existence of a one-parameter {\em helical symmetry}
$\chi_\lambda:\mathcal{M}\rightarrow \mathcal{M}$ \cite{FriUryShi02}
of the spacetime whose infinitesimal generator will be denoted by
$h^\mu$ ---that is,  $\frac{d\chi^\mu_\lambda}{d\lambda}= h^\mu$. Near the
binary system, this {\em helical Killing vector} $h^\mu$ is timelike.
Sufficiently {\em far} away $h^\mu$ becomes spacelike
\cite{FriUryShi02}, but a number $T>0$ exists such that the {\em
  separation} between a given point $p\in \mathcal{M}$ and its image
by the isometry flow $\chi_T(p)$, is timelike ---see also
\cite{Gou07a} for details.  The integral lines associated with $h^\mu$
are {\em helices}.

Helical symmetry is exact in theories with no gravitational radiation,
like Newtonian gravity, (second order) post-Newtonian gravity and
the Isenberg-Wilson-Mathews approximation to GR
\cite{Ise78,IseNes80,WilMat89}.  Helical symmetry can be exact in full
GR for non-axisymmetric systems with standing gravitational waves
\cite{BlaDet92,Det94}, although the spacetime cannot be asymptotically
flat, in the sense that the ADM mass cannot be defined \cite{GibSte83}
---more precisely, it cannot have a smooth null infinity if there is
no additional stationary Killing vector close to null infinity.
Helical Killing vectors have been used in the numerical literature to
model slow-motion adiabatic configurations of binary systems
\cite{WheRom99,WheKriPri00,WheBeeLan02,Price04}. This leads to the
study of mixed-type PDEs.  The {\em light cylinder}, characterised by
the null character of the helical vector $h^\mu$, separates an inner
domain where the PDE is elliptic from an outer one where it becomes
hyperbolic ---cf. also the so-called {\em periodic standing-wave}
approximation \cite{AndBeeBli04,
  BroOwePri05,BeeBroPri06,LauPri07,BeeBroHer07} and other recent
numerical works \cite{FriUry06,YosBroRea06}.  Further theoretical
developments making use of the helical Killing symmetry are given in
\cite{Kle04} and \cite{BeiHeiSch07}.  In particular, the work by Klein
---cf. his IHP talk \cite{IHP}--- aims at setting a consistent
framework for numerics by taking full advantage of the presence of a
Killing vector. He studies the EFE in the presence of a helical
Killing vector for a vacuum spacetime with two disconnected Killing
horizons of spherical topology. The use of a projection formalism on
the space of orbits of the Killing vector, permits then to cast the
problem in terms of the 3-dimensional gravity equations with a
$SL(2,\mathbb{R})/SO(1,1)$ sigma model as material source.

We must briefly comment on current constructions of binary black hole
initial data.  In the context of the XCTS construction ---see section
\ref{s:CTS}--- the choice of the evolution vector as an (approximate)
helical Killing vector, i.e. $t^\mu = h^\mu$, has been used in the
literature for {\em motivating} the Ansatz
$\dot{\tilde{\gamma}}_{\mu\nu}=0=\dot{K}$ for the free XCTS data.  The
helical Killing vector idea was applied for the first time to the
construction of binary black hole initial data in
\cite{GouGraBon02,GraGouBon02} and has subsequently been used in
\cite{CooPfe04,Ans05,Ans06,CauCooGri06}. All these data make use of an
excision technique to deal with the singularity ---see section
\ref{s:moving_punctures}.  Problems arise when combining XCTS and
helical symmetry with a puncture approach to initial data, as shown in
\cite{HanEvaCooBau03}. For this reason, punctured initial data make
use of the {\em extrinsic curvature} approach ---see section
\ref{s:conformal_Ansatz}--- which turns out to be difficult to
reconcile with a helical symmetry idea.  However, due probably to
their relatively large initial separations, linear superpositions of
two Bowen-York solutions ---firstly used in
\cite{Bau00,BakCamLouTak02,AnsBruTic04}--- have led to good results in
recent punctured binary black evolutions in
\cite{BakCenCho06,BakCenCho06b,MetBakKop06} and
\cite{CamLouMar06,CamLouZlo06b,CamLouZlo06c,CamLouZlo06d}.

\subsubsection{On the issue of moving punctures.}
\label{s:moving_punctures}
A key issue in numerical black hole simulations is the manner in which
black holes are modelled in the calculations. The two main techniques
employed are: {\em excision}, where a spatial neighbourhood of the
singularity is removed from the numerical grid on the initial
hypersurface and then subsequently, also on each spacelike
hypersurface constructed during the evolution
\cite{Tho87,SeiSue92,AnnDauMas95,Pre05,SpeKelLag05,SchPfeLin06,SziPolRez07}
---see also \cite{Mal96,Mal97} for particular analyses of global existence in this setting; 
and {\em punctures}, where one
begins with {\em punctured} initial data in the sense discussed in
section \ref{s:topology} and then evolves the data including the
singular point ---see for example
\cite{BruGonHan06,CamLouMar06,HerHinSho07,Spe07,GonSpeBru07,CamLouZlo07,BakMcWvMe07,ThoDiePol07,TicMar07}.
A third alternative replaces the singular black hole interior at each
spacelike slice by a regular one, leading to the idea of {\em stuffed
  black holes} \cite{ArbBonCar98,ArbBonMas99} that has been recently
brought back ---see \cite{BroSarSch07,EtiFabLiu07}. All these approaches
---excision, punctures and stuffing--- rely on the intuitive idea that
no information escapes from the black hole interior, and thus, assume
from the onset some sort of Cosmic Censorship. In particular, a
detailed analysis of this last ---non-trivial--- point has been
undertaken in \cite{BroSarSch07}.

There has been a recent interest in understanding some geometric
aspects of the moving punctures picture ---see e.g.
\cite{HanHusBru06,HanHusPol07,ThoDiePol07,Bro07}. However there seems
to be no fully consensed view on why the puncture method works, and
what happens with the punctures during the evolution. In particular,
it is of interest to know if they still represent a compactified
infinity. More importantly, it is crucial to know whether the method
relies on numerical errors near an under-resolved puncture, in which
case it could fail when probed at higher resolutions or if the
evolutions are let to run much longer than up to now.

Assuming the \emph{establishment point of view} on black holes, the
asymptotic state of the evolution will be described in some sense by a
Kerr spacetime. So, after the system has evolved long enough it is
natural to expect the existence of an \emph{approximate stationary
  Killing vector} that could be use to drive the evolution to an
eventually stationary slice.  A crucial element in the evolution of
punctured data is the use of the so-called \emph{symmetry-seeking}
choices of lapse and shift ---the {\em 1+log} lapse and the Gamma
freezing shift \cite{AlcBruDie03,MetBakKop06,GunMar06b}. A
symmetry-seeking gauge is one that tends to align itself with this
approximate Killing vector as the evolution proceeds. In this context,
for a stationary slice it is understood a hypersurface such that for
the chosen gauge-drivers, the evolution vector is parallel to the
stationary Killing vector ---so that the evolution of the spacetime
effectively freezes. The discussion of stationary slices in the
Schwarzschild spacetime has been elaborated further in
\cite{EstWahChr73,Rei73,HanHusBru06,HanHusPol07,BauNac07}.  Clearly,
it is of interest to carry out an analogous analysis for the Kerr
spacetime.

The \emph{symmetry-seeking} nature of the gauges used in the evolution
of dynamical black hole spacetimes is supported by numerical evidence.
According to this evidence, the evolution using symmetry-seeking
gauges seems to pile \emph{pointwise} in \emph{almost stationary
  slices} of the gauge.  It would be of great interest, both practical
and theoretical to have a rigorous analysis of this symmetry
seeking-behaviour of particular gauges. The resolution of the problem
of the non-linear stability of the Kerr spacetime is likely to clarify
this issue\footnote{It is perhaps of interest to note that in
  \cite{GarVal07} a particular prescription for a lapse and a shift
  ---in terms of 3+1 quantities--- has been calculated such that if
  the spacetime is the Schwarzschild solution, then these lapse and
  shift coincide with those implied by the timelike Killing vector. If
  one is analysing, say, the evolution of a head-on collision of black
  holes where one knows that the asymptotic state will be the
  Schwarzschild solution, then it is natural to expect these lapse and
  shift ---which can be evaluated in any spacetime--- to be also
  symmetry seeking. This particular analysis draws ideas from the
  invariant characterisation of spacetimes which has been extensively
  developed in the study of exact solutions to the Einstein field
  equations ---see e.g.  \cite{SKMHH}.}.

General results on the topology of hypersurfaces in asymptotically
flat spacetimes ---see e.g.  \cite{Gan75,Gan76}--- preclude the change
of topology during the evolution, unless the slices touch the
singularity ---see also the discussion in \cite{MisThoWhe}, section
31.6. Numerical evolutions ---together with analytical
considerations--- of punctured Schwarzschild initial data using
symmetry-seeking gauges \cite{HanHusBru06,HanHusPol07} show indeed
that the solution approaches {\em pointwise} a stationary slice that
does not hit the physical singularity. Most importantly, the limiting
slice does not reach the inner asymptotically flat spatial infinity,
but rather ends on a cylinder of finite areal radius whose throat has
an infinite proper distance as measured from the apparent horizon.
This is interpreted as a change of the geometry of the puncture along
the evolution ---from a $1/r$ behaviour of the conformal factor
near the puncture to $1/\sqrt{r}$--- that suggests the use of
\emph{asymptotically cylindrical} data to represent black holes ---see
also the discussion in the IHP talks by Hannam and O'Murchadha.  These
results have also been discussed in \cite{Bro07}, where it is pointed
out that the puncture actually remains at the inner spacelike infinity
along the whole evolution and, in particular, the evolution using
positive lapse is never stationary in the limit of infinite
resolution.  Both apparently contradictory interpretations can be
reconciled in terms of the differences between uniform and pointwise
convergence: the limit to the final cylindrical stationary slice seems
to be pointwise, but not uniform.  In actual numerical simulations,
the punctured region is under-resolved and there is no practical need
to distinguish between both kinds of convergence.  In particular, this
lack of resolution is not likely to cause problems for finite
difference codes at any reasonable ---practically realistic---
resolution.  In this interpretation, the moving puncture evolution can
be seen as an effective {\em natural excision} method \cite{Bro07}.
It is expected that the mechanisms operating in the case of
simulations for Schwarzschildean initial data are in essence the same
ones working in numerical evolutions of dynamical black hole
spacetimes ---see comments in \cite{Bro07} about the presence of spin
and momentum.

\subsubsection{Discussion.}
The recent intense activity in numerical evolutions of binary black
holes has meant a very rapid advance in the understanding of the
physics of binary black holes ---with a particular emphasis on
astrophysical applications. Although more systematic studies are needed, 
a large number of results are already available, for
example concerning
the effect that different configurations of physical
parameters ---such as mass ratios, spins or orbit eccentricities--- 
have on the final state of the resulting black hole,
in particular on the recoil
velocities of the final black hole.
Again we refer to \cite{Pre07} for a bibliographic account of these
advances ---cf. also 
the IHP talks by Lousto, Lindblom and Van Meter.

In what concerns astrophysical applications,
one should mention
the complementarity of numerical approaches with others of analytic nature.
On the one hand, given the computational cost of the numerical simulations
it is natural to explore analytic or semi-analytical methods to cover the
whole parameter space of the problem. On the other hand, analytical
approximations often offer a (quick) hint on the physics of the problem.
The need of such a synergy between numerical and analytical efforts
was advocated by Damour and Blanchet in their respective IHP talks 
\cite{IHP}, where different post-Newtonian approaches were 
confronted with numerical results \cite{DamNag07,BlaQW05}. 
Black hole perturbation theory offers
a further instance of potentially fruitful combination 
of analytical and numerical approaches, for example in the analysis
of the extreme mass ratio
case ---cf. Nagar's IHP talk \cite{IHP}.

Astrophysical aspects of the binary black hole problem offer plenty of
room for the collaboration between analysts and numerical relativists.
In addition, the very success of the simulations already offer {\em
  some indirect} evidence of geometric issues like Cosmic Censorship
or Kerr stability.  However, the numerical control of this non-linear
problem opens an outstanding possibility of gaining insights into some
of the key geometric and conceptual problems of GR, if specific
efforts are geared in this direction.  Geometric ideas and results,
such as Cosmic Censorship or black hole uniqueness theorems, have
built a solid conceptual framework for the study of gravitational
collapse.  In turn, it is to be expected that strong numeric tests to
some of these geometric ideas could be devised, offering a window to
the assessment of non-linear features in GR. In current simulations,
non-linear effects in the binary black hole have resulted unexpectedly
mild. But, as discussed for example in Lehner's IHP talk \cite{IHP},
``unchartered trails'' remain in the road and much should be learnt
from studies of generic configurations ---cf. in this sense
\cite{LouZlo07,CamLouZlo07b} where multi-black hole encounters are
studied.

\section{Miscellaneous topics} \label{s:miscellaneous}

There are some important topics in the relation
between geometry and numerics that have not been 
discussed in previous sections. 
Here we briefly present ---non-exhaustively--- some of those aspects 
and lines of research.

\subsection{Critical collapse}
\label{s:critical}
The study of critical phenomena\footnote{The content of his section is essentially due to J.M. Mart\'{\i}n-Garc\'{\i}a. We are thankful for his enthusiastic input.} in gravitational collapse is one of
the paradigms of the interaction between
numerical and mathematical relativity: a type of phenomena which was
discovered by means of numerical experiments \cite{Chopt92} which then, in turn, have
been understood by analytical methods ---see \cite{GunMar07,Ber02} for 
a review.
In 1992 Choptuik \cite{Chopt92} found that it is possible to form arbitrarily small
black holes in the process of gravitational collapse of a spherical
scalar field, by fine tuning any parameter which affects the
self-gravity of the initial configuration. Before the black hole is formed the
spacetime becomes selfsimilar and universal, forgetting the initial
condition except for a single length scale which determines the mass
of the black hole to be formed. This interesting behaviour has been
found in many other matter systems, and perturbative arguments suggest
that it could be also present without spherical symmetry, being
therefore intrinsic to the strong regime of GR dynamics. Finally,
perfect fine tuning generates a naked singularity with infinite
curvature, visible to observers at infinity, with important
consequences for Cosmic Censorship and Quantum Gravity.  These
{\em critical phenomena in gravitational collapse} were completely
unexpected before 1992 and are now considered the best example of
how numerical relativity can contribute new physics to GR  ---see Aichelburg's and Harada's IHP talks for examples of ongoing research.  
A posteriori, they can be understood as the discovery of
codimension 1 exact solutions in the infinite-dimensional phase space
of General Relativity ---Minkowski, black holes and stars begin global
attractors of codimension $0$--- opening a new way of interpreting GR
from the point of view of the theory of dynamical systems.

\subsection{Quasi-local physical parameters}
\label{s:quasi-local_param}
The determination of physical parameters in a finite region of spacetime
is of clear practical importance in numerical 
implementations. More generally, it is of conceptual relevance in the
theory as a whole and, more concretely, in some particular geometric constructions such
as certain geometric flows  \cite{Gib97,Gib99,BraChr04,HuiIlm01,BraHayMar06} 
or in the study of geometric inequalities ---cf. section \ref{s:geom_ineq}.
The determination of the amount of energy in a compact domain is a classical
problem in GR  and an exhaustive list of proposals regarding this goes beyond the scope
of this article ---see \cite{Sza04}. 
The relevance of this subject as a boost for the interaction 
between geometry and numerics is illustrated, for example, in 
the use of quasi-local dynamical trapping horizons
---section \ref{s:quasi_bh}--- for the extraction of physics in current 
numerical simulations. We highlight the attempts
of defining a quasi-local angular momentum by means 
of an integral on a closed surface ${\cal S}$ involving 
an {\em axial} vector $\phi^i$ defined on  ${\cal S}$
---e.g.
\cite{BroYor93,AshBeeLew01,Sza04,AshKri03,AshKri04,BooFai05,Gou05,Hay06,Sza06,Kor07}, 
see also the related work in \cite{Sza07}---
and in particular the recent efforts for 
deriving an unambiguous prescription for 
$\phi^i$ \cite{DreKriSho03,CooWhi07,Kor07,Sza06}. Note that the divergence-free 
character of $\phi^i$ in not guaranteed in all the latter schemes, a relevant
point if a slice-independent definition ---i.e. intrinsic to ${\cal S}$---
of the angular momentum is desired. In addition to determining the magnitude of
the spin, this vector $\phi^i$ can be used
to estimate the direction of the angular momentum vector ---cf. \cite{CamLouZlo07} for
preliminary results.

\subsection{Spacetime singularities and extensions of GR}
The study of the limits and extensions of GR offers stimulating
perspectives of research. One the one hand, numerical studies of
singularities \cite{Ber02,Poiss97,HodPir98,Gar04}  
---e.g. Cauchy horizon instabilities and
mass inflation near the Cauchy horizon, naked and
null singularities,  approach to spatial
singularities and analysis of BKL conjecture--- explore the internal consistency of
the theory and offer insight into new conceptual developments in it.  
Numerical implementations of higher dimensional spacetimes as
well as the coupling of the geometry to additional fundamental fields
---e.g. in Einstein-Yang-Mills theory--- probe extensions of the theory
that can be relevant in the quantum context.

\subsection{Some astrophysical considerations}
Ultimately motivated by the astrophysical study of relativistic
compact objects, this article has focused on isolated systems in GR.
Moreover, we have centred the discussion on black holes, since they
offer a particularly rich context for the dialogue between geometers and
numerical relativists.  This methodological choice could shadow the
extraordinary developments in astrophysical numerical relativity
regarding systems with matter.  Examples of the latter are the results
relative to compact objects, such as neutron or strange stars ---
notably the first computations of merger of binary neutron stars by
Shibata and Ury\=u \cite{ShiUry00}; see also Ury\=u's and
Saijo's IHP talks \cite{IHP}--- and other advances in relativistic
hydrodynamics \cite{AndCom07,Font03,MarMul03}. This constitutes
presumably an extremely
active area of research in next future.  Moreover, the study of
problems such as the Einstein-Vlasov system presents 
an analytic as well as numeric interest \cite{Andre05,AndRei06}.

\subsection{Geometrical spacetime discretisations.}
In section \ref{s:IVP}, initial value problem formulations of GR have been 
presented as a particularly powerful approach to the construction of generic 
spacetimes. A radically different approach
consists in adopting a formulation of GR which avoids the use of coordinates in the
spirit of the Regge calculus ---see e.g. \cite{Reg61, Gen02}. 
For a current programme of research in these lines, see \cite{Fra06,RicFraVog07}
---and also Frauendiener's IHP talk \cite{IHP}---  which is
based on Cartan's method of frames and the use of discrete differential
forms.

\subsection{Numerical techniques}
As we have commented in section \ref{s:evol_formalism}, a
comprehensive presentation of the evolution formalisms used in
numerical relativity should be complemented with a discussion of the
employed numerical techniques. Topics offering plenty of occasion for
the collaboration between analysts and numerical relativists are,
among others, adaptative mesh refinement, high order methods (cf. Tiglio's
IHP talk)  spectral methods \cite{GraNov07} ---including their 
application to evolutions \cite{HenAns08}--- and spectral elements (see
also Maday's IHP talk), finite elements (e.g. \cite{Holst01,SopLag05,SopSunLag05,Sop06,KorAksHol08,AksBerBon08}) 
finite volumes (e.g. \cite{AliBonBon07}) or advances in 
high-resolution shock capturing methods.

\subsection{Algebraic symbolic calculus}
The increase in computational capabilities and the development of
powerful new computer algebra systems has had a great impact in many
areas of GR. Many problems which for a long time seemed to be out
reach merely on the grounds of computational complexities are becoming
feasible. Algebraic symbolic methods were firstly used in GR as a tool
in the general research area of \emph{exact solutions}. Besides its
evident utility in the derivation of exact solutions, computer algebra
systems had a significant application in the \emph{metric equivalence
  problem} ---which consists in deciding whether two metrics given in
arbitrary coordinates are locally isometric to each other. The
equivalence of two metrics is a classical problem in Differential
Geometry, and a solution was given by Cartan \cite{Car46}. Early
discussions of the equivalence problem in GR can be found in
\cite{Bra65} ---see also \cite{Kar80a,Kar80b}.  A particular
implementation of these ideas is the system {\tt CLASSI} ---see
\cite{Ama02}. Considerations involving the equivalence problem can be
of utility in the comparison of numerical simulations.

In recent years, algebraic symbolic calculus has been increasingly
used as a systematic tool for analysing the analytic properties of
evolution systems ---e.g. in the construction of symmetric systems,
analysing characteristics, in the construction of propagation systems,
construction of asymptotic expansions and in perturbative analyses.
Among the systems explicitly constructed with this purpose in mind one
has {\tt xAct} \cite{xAct}.  Of relevance is also the use of computer
algebra systems for the automatic generation of computer codes like in
the case of {\tt Kranc} \cite{HusHinLec06}.

\section{Conclusions}
\label{s:conclusions}
The extraordinary results in the numerical evolution of black hole
binaries have had and will have a deep impact on the relation between
mathematical and numerical relativists. In particular, it forces a
reflection on the long term scientific objectives of the research of
both communities. On the one hand it questions the potential pay-offs
of certain lines of investigation, and on the other hand it offers the
possibility of addressing problems which for \emph{technological}
reasons were considered out of reach. In particular, theory will have
the unique opportunity to confront observation by means of the
accurate simulation of relativistic astrophysical systems.
Previsively, a big proportion of the numerical community will be
involved in this endeavour. Arguably, the implementation of the latter
will not require many geometrical insights, although the invariant
extraction of physical content would certainly benefit from it
---e.g. through the development of more efficient AH-finders, 
quasi-local characterisations of physical quantities, 
 or invariant algorithms for the extraction of gravitational
radiation.
However, if one wants to study the other fundamental aspect of the
theory, namely the structural ones ---exploiting the state-of-the-art
numerical possibilities--- then evidently a geometrical perspective is
fundamental. Even more, one could argue that the study of certain
crucial geometrical aspects of GR will require numerical approaches to
come to fruition.

In this review we have focused on the latter geometric aspects of numerical
relativity and also on genuinely mathematical results which, we
believe, are of relevance for numerical applications. There
is an intrinsic aesthetic appeal in the study of intrinsic geometric
aspects of GR, but the benefit goes much further. We have tried to
emphasise the role ---and necessity--- of analytical studies as
guarantors of the internal consistency of the theory.
Geometric-oriented lines research can and do offer general conceptual
frameworks in which physically motivated problems can be unambiguously
formulated. The gravitational collapse paradigm is an example of this.
On a second stage, it is also undeniable that Geometry also provides
powerful tools and insights into calculating things. Following the
premise that a crucial aspect of ``understanding a theory consists in understanding its solutions'',  \emph{geometrical numerical relativity} is one of the
most powerful tools available to study the space of solutions of GR.
Numerical GR can, potentially ---i.e. assuming enough computer
resources--- solve to any desired finite precision any well
formulated problem concerning a property of a specific solution
in a concrete problem. When the problem involves an infinite number
of solutions, then only the generic behaviour can be analysed.
This is however where numerical GR can be most useful, finding new
and unexpected results
\footnote{We acknowledge J. M. Mart\'{\i}n-Garc\'{\i}a for bringing
  out this point.}.

We finalise by naming a number of problems for which, we believe, the
interaction between numerical and mathematical relativists is of
particular relevance ---this, of course, notwithstanding the other
issues that have been suggested in the main text. Cosmic Censorship
remains in the eyes of most relativists the most important open
problem in classical GR. It is to be expected that numerical
investigations of global spacetimes will prove of utmost value in the
strive towards a proof of the conjectures. In the closely related
issue of the non-linear stability of the Kerr spacetime, numerical
investigations have already provided information about the decays of
various fields. This interaction is bound to become even closer in the
future. A related topic, the discussion of the robustness aspects
---with respect to changes in the initial data--- in the production of
gravitational radiation by isolated systems will require strong input
from numerics ---even to obtain a rigorous formulation of the problem.
The assessment of the physical relevance of certain aspects of the
characterisation of isolated systems through the notion of asymptotic
simplicity ---like the peeling behaviour--- will also require close
numerical examination. Something similar can be said about the
relevance of the construction of initial data sets by means of gluing.
Close interaction between analytics and numerics will be required in
the study of dynamical trapping horizons and their asymptotic
behaviour close to the event horizon and in the relation between the
global and local characterisation of the black hole notion
---extension of the trapping boundary to the event horizon. Finally, a
study of the solution space of GR ---in the spirit of the theory of
dynamical systems--- using ideas developed in the study of critical
phenomena should also be based on a close interaction between numerics
and geometry.

\section*{Acknowledgements}
We would like to thank all participants of the ``From Geometry to
Numerics'' workshop, part of the ``General Relativity Trimester'' at
the Institut Henri Poincar\'e for the effort put in preparing their
presentations. We also thank T. Damour and N. Deruelle for having organised
the General Relativity Trimester. 
 We are particularly indebted to J.M. Mart\'\i
n-Garc\'\i a and A. Zengino\u{g}lu for their contributions to sections
\ref{s:critical} and \ref{s:conformal_equations}, respectively. All misunderstandings or omissions in
these sections are, nevertheless, our responsibility. We also would
like to thank V. Aldaya, M. Ansorg, C. Barcel\' o, N. Bishop,
S. Bonazzola, C. Bona, I. Cordero-Carri\'on, S. Dain,
P. Grandcl\'ement, B. Krishnan, S. Husa, J.M. Ib\'a\~nez,  L. Lehner, 
E. Malec,
M. Mars, G.
Mena-Marug\'an, V. Moncrief, J. Novak, T. Pawlowski, M. S\'anchez, 
J. Seiler, J.M.M. Senovilla, L. Szabados, 
N. Vasset and D. Walsh
for very valuable discussions and comments.  JLJ acknowledges the support
of the Marie Curie contract MERG-CT-2006-043501 in the 6th European Community
Framework Program and wish to thank the School of Mathematical Sciences
of the Queen Mary for its hospitality.
JAVK is supported by an EPSRC Advanced
Research Fellowship and wish to thank the Instituto de
Astrof\'\i sica de Andaluc\'\i a (CSIC) for its hospitality.
EG acknowledges support from the ANR grant 06-2-134423 
\emph{M\'ethodes math\'ematiques pour la relativit\'e g\'en\'erale}.


\section*{References} 

\end{document}